\documentclass[prd, onecolumn, floatfix, letterpaper, nofootinbib, amsmath, amssymb, superscriptaddress,showpacs]{revtex4}
\usepackage{caption}
\usepackage{graphicx}
\usepackage{epsfig}
\usepackage{bm}
\usepackage{amsfonts}
\usepackage{comment}
\usepackage{tikz}
\usetikzlibrary{shapes,arrows}

\usepackage{color}

\newbox\pippobox

\def\be{\begin{equation}}
\def\ee{\end{equation}}
\def\ba{\begin{eqnarray}}
\def\ea{\end{eqnarray}}
\newcommand {\lla} {\ {\raise-.5ex\hbox{$\buildrel<\over\sim$}}\ }

\usepackage[T1]{fontenc}
\usepackage[latin1]{inputenc}
\usepackage{graphicx}
\usepackage[english]{babel}
\usepackage{amsmath}
\usepackage{amssymb}
\usepackage{amsfonts}

\def\e{\mathrm{e}}

\def\be{\begin{equation}}
\def\ee{\end{equation}}
\def\bea{\begin{eqnarray}}
\def\eea{\end{eqnarray}}

\newcommand{\lsim}{\lesssim}
\def\spose#1{\hbox to 0pt{#1\hss}}

\def\lta{\mathrel{\spose{\lower 3pt\hbox{$\mathchar"218$}}
     \raise 2.0pt\hbox{$\mathchar"13C$}}}
\def\gta{\mathrel{\spose{\lower 3pt\hbox{$\mathchar"218$}}
     \raise 2.0pt\hbox{$\mathchar"13E$}}}

\newcommand{\Mp}{M_{_\mathrm{Pl}}}

\begin{document}

\title{Backreaction and Stochastic Effects in Single Field Inflation}

\author{ Laurence Perreault Levasseur} 
\email{lp375@cam.ac.uk}
\affiliation{DAMTP, University of Cambridge, Cambridge, CB3 0WA, United Kingdom}
\affiliation{Kavli Institute for Particle Astrophysics and Cosmology (KIPAC), 
Menlo Park, CA 94025, USA}
\author{Evan McDonough} 
\email{evanmc@physics.mcgill.ca}
\affiliation{Department of Physics, McGill University, Montr\'eal, QC H3A 2T8, Canada}

\begin{abstract}
The formalism of stochastic inflation is a powerful tool for analyzing the backreaction of cosmological perturbations, and making precise predictions for inflationary observables. We demonstrate this with the simple $m^2 \phi^2$ model of inflation, wherein we obtain an effective field theory for IR modes of the inflaton, which remains coupled to UV modes through a classical noise. We compute slow-roll corrections to the evolution of UV modes (i.e. quantum fluctuations), and track this effect from the UV theory to the IR theory, where it manifests as a correction to the classical noise. We compute the stochastic correction to the spectral index of primordial perturbations, finding a small effect, and discuss models in which this effect can become large. We extend our analysis to tensor modes, and demonstrate that the stochastic approach allows us to recover the standard tensor tilt $n_T$, plus corrections.
\end{abstract}

\maketitle
\vspace{-0.5cm}
\tableofcontents

\section{Introduction}
Inflation has been tremendously successful in explaining the physics of the very early Universe. It was the first compelling cosmological model to provide a causal mechanism for generating fluctuations on cosmological scales, and it predicted that their spectrum should be almost scale invariant, with small deviations from scale invariance that can be traced back to the precise microphysics of inflation~\cite{Mukhanov:1981xt,Mukhanov:1982nu,Bardeen:1983qw}. These predictions provide a way of connecting theoretical physics to observational cosmology; this has been a very fruitful venture, as has lead to particle-physics based models of inflation \cite{Lyth:1998xn}, inflation in supergravity \cite{Kallosh:2014vja}, and string inflation \cite{Baumann:2014nda}, to name a few.  There is still much to be learned from the CMB, and if the large tensor-to-scalar ratio of \cite{Ade:2014xna} is a hint of good things to come, then the CMB may yet give us an unprecedented opportunity to test models of inflation and quantum gravity. 

With the ever-increasing precision of experiments probing the CMB, for example~\cite{planck}, it becomes imperative to develop self-consistent methods of calculation for inflationary predictions. The formalism of stochastic inflation is a promising avenue in this direction. It allows for the constant renormalization of background dynamics and in this way circumvents one of the main difficulties of traditional methods: backreaction~\cite{Mukhanov:1996ak, Brandenberger:2002sk, Geshnizjani2005, Martineau:2005aa, 2010GReGr..42.1399K, Levasseur2010}. This is achieved by separating the dynamics of long, classical wavelengths from short, quantum fluctuation-dominated wavelengths, and studying the interplay of the two sectors.  The stochastic formalism then allows for the resummation of corrections to the background dynamics as modes of fluctuations are stretched from the quantum regime into the coarse-grained effective theory. 

The resulting theory describes the effective classical dynamics of a large-scale gravitational system, in the presence of a `bath' where all the quantum fluctuations are collected in a classical noise term, through a set of Langevin equations. As required by the fluctuation-dissipation theorem, this noise term comes hand in hand with a dissipation term, which in turn allows for irreversibility and approach to equilibrium. The effective theory therefore belongs to a new class of non-Hamiltonian theories \cite{Burgess:2014eoa}, which have not been studied much so far in the context of cosmology. (However, see~\cite{Berera2008} and references therein in the context of warm inflation.)

Stochastic inflation has a long history. Originally proposed by Starobinsky \cite{Starobinsky1984, Starobinsky1986}, stochastic inflation as studied in the early work of \cite{Vilenkin1983a, Vilenkin1983b, Bardeen1986, Rey1987, Goncharov:1987, Sasaki:1987gy, Nakao:1988yi, Nambu:1988je, Nambu:1989uf, Salopek1990, Salopek1991, Habib1992, Linde:1993xx, Starobinsky1994} was a simple way to include quantum effects into inflation. The idea was this: quantum fluctuations are generated deep inside the horizon and, at zeroth order in slow-roll, evolve as quantum fields on a fixed de Sitter background. The quantum modes grow and exit the horizon. Doing so, due to their random phase, they provide a kick of a random amplitude to the long-wavelength physics. It follows that the quantum modes act as a source for the classical background, and the physics of this source is probabilistic in nature. More precisely, stochastic inflation provided an `educated guess' that this source should be white noise. The physics of slow-roll inflation can then be studied as per the usual treatment, with the noise included as a source in the equation of motion for the classical (long-wavelength) field.

Stochastic inflation was put on a more solid footing by \cite{Morikawa:1989xz} and \cite{Hosoya:1988yz}, where the equations of motion for stochastic inflation were derived from a path integral~\cite{Calzetta:1993qe, Calzetta1995, Calzetta1997, Matacz1997, Calzetta1999, Matarrese2004, Franco2010}. Given these equations of motion, the vast majority of modern applications of stochastic inflation take the same approach as Starobinsky: calculate the variance of quantum modes in a pure de Sitter background, include this as white noise in the Klein-Gordon equation for long-wavelength modes, and study slow-roll inflation in the presence of this white noise (see, e.g., \cite{Afshordi2001, Geshnizjani2005a,Tsamis2005, Martin2006, Kunze:2006tu, Finelli:2008zg, Finelli:2010sh, Clesse:2010iz, Martin2011, Weenink2011}). However, this method misses a key element of the physics: as pointed out in \cite{Boyanovsky1995} the short-wavelength and long-wavelength physics are coupled. Namely, the quantum modes do not evolve on a pure de Sitter background, but rather on a background that is both slow-roll and stochastically corrected. In terms of the path integral, the coupling of the two sectors (long-wavelength, or `coarse-grained' fields, and short-wavelength, or `bath' fields) manifests itself as loop diagrams calculated in the Schwinger-Keldysh `in-in' formalism of quantum field theory, which has become widely applied in cosmology since \cite{Weinberg:2005vy, Maldacena:2002vr}, after the early work of \cite{Calzetta1987, Jordan1986} (however, see \cite{Altland2010, Calzetta2008} and references within for an introduction in the context of out-of-equilibrium QFT and open systems). This approach was developed in \cite{Levasseur:2013ffa}, where it was dubbed the `recursive formalism of stochastic inflation'.

Cosmological perturbations have also been studied in the context of stochastic inflation, see for example \cite{Kunze:2006tu} and \cite{Fujita:2013cna, Fujita:2014tja}. We will use a method inspired from the approach used in \cite{Kunze:2006tu}, with some modifications that will be discussed in section VI. An alternative, and relatively recent, proposal \cite{Fujita:2013cna, Fujita:2014tja} is to apply the $\delta N$ formalism to stochastic inflation. This makes intuitive sense: the $\delta N$ formalism can be qualitatively understood as a `separate universe approach', and one would not expect a local noise to invalidate this approach. This approached will also be touched upon in section VI.

In \cite{Levasseur:2013tja}, the recursive formalism was applied to hybrid inflation~\cite{Linde:1993cn}. In this scenario, the spectral index is strongly dependent on the duration of the `waterfall phase' of inflation~\cite{Clesse2009}, where the field dominating the energy density of the Universe during inflation becomes tachyonic and `waterfalls' down the side of the potential. This generates a red tilt, provided that the waterfall phase lasts for a suitable number of $e$-folds. It was found in \cite{Levasseur:2013tja} that the recursive corrections caused the tilt of the inflaton perturbations to become bluer in the valley, while also causing the waterfall phase to end earlier than otherwise expected, making a red tilt much more difficult, if not impossible, to achieve.

In the present paper we have more modest goals, that is, to study recursive stochastic effects in single field inflation, both analytically and numerically, in particular the simple $m^2 \phi^2$ model~\cite{Linde1983177, Linde:1993xx, Linde:1994yf}, away from the regime of eternal chaotic inflation~\cite{Linde1986eternalinflation, Linde1986395, Guth:2007ng}.\footnote{ See also \cite{Li:2007uc, Kohli:2014ala, Feng:2010ya, Qiu:2011zr} and references therein for existing studies of stochastic eternal inflation.} We find that the recursive approach gives corrections to quantum modes that could not be deduced from naively including slow-roll effects alone. We then study the effect of this on long-wavelength perturbations, and calculate the power spectrum of primordial perturbations. We extend this approach to include tensor perturbations, and discuss the effect of couplings to heavy fields.

The outline is as follows: in section II we outline the usual approach to stochastic inflation and review the recursive formalism. In section III we calculate the classical noise induced by quantum fluctuations on a classical background which is zeroth order in slow-roll. In section IV we study the effect of this noise by computing the stochastic (and slow-roll) corrected classical background, and continue in section V to compute the backreaction on the quantum modes. We then use this in section VI to compute the backreaction on IR modes and the spectrum of curvature perturbations. We extend this to a class of simple multifield models in section VII, and to tensor modes in section VIII. We conclude and discuss our results in section IX.

\section{Stochastic Inflation: Basic Setup and Recursive Strategy}

Let us first consider the action of a single scalar field in a fixed background. The matter part of the action is given by:
\begin{equation}
S_M = \int d^4{x} \sqrt{-g} \left(-\frac{1}{2} \partial_\mu \Phi \partial^\mu \Phi - V(\Phi) \right) \, ,
\end{equation}
which leads to the equation of motion
\begin{equation}
- \Box \Phi + V, _{\Phi} = 0 \, ,
\end{equation}
\begin{equation}
\Box = -\partial_{tt} - 3 H \partial _{t} + \frac{\nabla^2}{a^2} \, .
\end{equation}
In the present paper, we will more specifically be interested in the chaotic potential
\be
V(\Phi) = \frac{1}{2} m^2 \Phi^2\, .
\ee
Moreover, to ensure that we remain away from the eternal inflation regime throughout our analysis, we impose the condition $m\Phi_0^2/\Mp^3\ll 8\pi$ throughout this paper~\cite{Guth:2007ng}, where $\Phi_0$ is the initial value of the inflaton at the beginning of inflation.

The starting point of stochastic inflation is to split the field $\Phi$ into long-wavelength modes $\phi_c$ (c for classical), and short-wavelength modes $\phi_q$ (q for quantum). Note that both $\phi_q$ and $\phi_c$ are quantum fields in nature;  $\phi_c$ technically corresponds to a quantum averaged field, coarse grained on a radius of constant physical size. We choose this coarse-graining scale to be the scale at which quantum fields undergo squeezing, i.e. the Hubble scale., at which point the commutators of the fields and their derivatives scale as $k/a$ and are therefore exponentially suppressed (see \cite{Calzetta1987, Habib1992, Albrecht1992, Calzetta1995, Hu1995, Polarski1995, Calzetta1997, Matacz1997, Kiefer1998, Kiefer2006, Kiefer2007, Franco2010, Weenink2011, Burgess:2014eoa} and references therein concerning the topics of quantum average versus classical fields, decoherence, and the conditions required for classicalization). 

The splitting into $\phi_c$ and $\phi_q$ is defined by
\begin{equation}
\Phi = \phi_c + \phi_q \, ,
\end{equation}
\begin{equation}
\label{phiqmodes}
\phi_q = \int \frac{\mathrm{d}^3k}{(2\pi)^3}  W(k,t) \hat\Phi_k e^{-i k \cdot x} \, ,
\end{equation}
where the $\hat\Phi_k= \phi_k \hat a_k+\phi_{-k}^*\hat a^\dagger_{-k}$ is the mode expansion of the quantum fields in terms of creation and annihilation operators, and $W(k,t)$ is a time-dependent window function. The window function acts to project onto $\phi_q$ only the modes with a comoving wavenumber somewhat larger than the physical Hubble scale. To be precise, we take the filtering scale to be large enough to make sure that all modes that are part of $\phi_c$ have undergone squeezing and classicalization. To see this splitting at the level of the equation of motion, we can Taylor expand the equation of motion\footnote{In a dynamical spacetime $H$ depends on the \emph{full} quantum field $\Phi$, and hence the extension of this heuristic argument to realistic inflationary setups is slightly more involved, although conceptually the derivation is identical. } about $\Phi=\phi_c$,
\begin{equation}
\label{eq:firstexpansion}
	 -\Box \phi_c + V_{,\Phi} (\phi_c) + \left[ - \Box \phi_q + V_{,\Phi \Phi} (\phi_c) \phi_q \right] = - \frac{1}{2}V_{,\Phi \Phi \Phi} \phi_q \phi_q  + ... \, ,
\end{equation}
where our perturbation variable has been chosen to be the number of quantum fields (which can be seen from the path integral formulation to be equivalent to counting powers of $\hbar$ in a Schwinger-Keldysh loop expansion).

Given that the coarse-graining radius is chosen to correspond to the classicalization radius, the quantum-averaged field $\phi_c$ corresponds to an effective classical field which we call $\tilde\phi_c$, endowed with a probability density function (PDF). This effective classical field $\tilde\phi_c$ allows us to treat collectively all realizations of the universe with consistent histories. Its PDF gives different probabilistic weights to classical realizations coming from different sets of random phases of the mode functions as they successively cross the Hubble radius and freeze. The PDF allows for a notion of ensemble average, which is equal to the quantum expectation value provided the ergodic hypothesis is satisfied. This point is further clarified in \cite{Tsamis:2005hd} and discussed in more details in \cite{Levasseur:2013ffa}.

It follows that alongside the system $\{ \phi_c$ , $\phi_q \}$, \{coarse-grained quantum field , small scale quantum fluctuations\}, we can write a corresponding classical, probabilistic system consisting of $\tilde\phi_c$ and a set of classical Gaussian noises $\xi _{1,2}$ modeling the effects of the incoming modes of $\phi_q$ joining the coarse-grained theory. Using the definition (\ref{phiqmodes}) to rewrite the $\phi_q$ in square brackets in terms of their linear mode expansion, as well as the fact that the linearized mode functions $\phi_k$ satisfy their linearized equation of motion, equation (\ref{eq:firstexpansion}) can be rewritten as
\begin{equation}
\label{eq:phic}
	-\Box \tilde\phi_c + V_{,\Phi} (\tilde\phi_c) = 3H\xi_{1} +\dot\xi_1-\xi_2~~+ \left(- \frac{1}{2} V_{,\Phi \Phi \Phi} \phi_q \phi_q  + ...\right) \, ,
\end{equation}
where the only surviving terms in the square brackets (i.e. the ones containing at least one time derivative acting on the time-dependent window function $W(k,t)$) have been defined as the classical noise. A simple calculation, again using only equation (\ref{phiqmodes}), reveals that the noise terms are drawn from a random Gaussian probability distribution given by
\begin{equation}
\label{eq:masterPDF}
		\mathcal{P}\left[\xi_1, \xi_2 \right]= \exp\left\lbrace -\frac{1}{2}\int d^4xd^4x' [\xi_1 (x)\, \xi_2(x)] {\bf A}^{-1}(x,x')\left[\begin{array}{c}\xi_1(x')\\  \xi_2(x') \end{array}\right] \right\rbrace\, ,
\end{equation}
and we have defined, letting $r=|{\bf x}-{\bf x'}|$, the matrix $A$ to have components given by
\be
\label{themonstercorrelationmatrix}
{\bf A}^{i,j}(x, x')= \int \frac{d k}{2\pi^2}  \frac{\sin (kr)}{kr}  \partial_{t}W\left(k, t\right)\partial_{t'}W\left(k, t'\right)\mathrm{Re}\left[ {\bf M}^{i,j}(k, t, t')\right]\, ,
\ee
with
\be
\label{theMmatrices}
	 {\bf M}^{i,j}({\bf k}, t, t')=\left(\begin{array}{cc} \phi_{\bf k}( t) \phi^*_{\bf k}(t')  & \phi_{\bf k}(t) \dot\phi^*_{\bf k}(t') \\  \dot\phi_{\bf k}(t) \phi^* _{\bf k}(t')& \dot \phi_{\bf k}(t) \dot\phi^*_{\bf k}(t')   \end{array}\right)\, ,
\ee
where ${\bf k}$ in the above equation is larger than the coarse-graining scale. The terms written in parenthesis on the right-hand side (r.h.s.) of (\ref{eq:phic}), which are the only ones still containing quantum fields, can also be rewritten in terms of classical noise terms,\footnote{In fact, equation (\ref{eq:phic})  is inconsistent \emph{unless} this is done, since as written they are a quantum contribution to a classical equation of motion.} as was done in \cite{Levasseur:2013ffa}. However, for the case of a quadratic potential, which we shall consider here, these higher-order terms vanish and hence will not contribute to equation (\ref{eq:phic}). To emphasize the split between quantum and classical: the modes $\phi_q$ are quantum, while the noises $\xi_{1,2}$ are \emph{classical}, as the noise terms appearing in equation  (\ref{eq:phic}) are evaluated at the moment the modes $\phi_q$ exit the horizon and give a `kick' to the IR (classical) theory. 

The variance of $\xi_1$  and $\xi_2$ can be read directly from this definition, by equating ensemble averages and quantum expectation values under the ergodicity assumption. To solve for the stochastic background, it is necessary to solve simultaneously for the linear mode function of the bath field, which satisfies the equation:
\be
\label{modefunctioneom}
	\left(\partial_t^2+3H\partial_t-\frac{k^2}{a^2}+m^2\right)\phi_{\bf{k}}(t)=0 \, ,
\ee
where the wavenumber ${\bf k}$ is larger than the coarse-graining scale, i.e. for wavelengths smaller than the coarse-graining radius.
In what follows, we will be interested in solving the classical system $\{\tilde\phi_c, \xi _q\}$ perturbatively and will not look any further at the quantum averaged field $\phi_c$. We therefore drop the tilde for the sake of simplicity, and from now on by $\phi_c$ we mean the classical, stochastic coarse-grained field. 

This is not an easy system to solve: the coarse-grained field $\phi_c$, which obeys (\ref{eq:phic}), depends on the amplitude and statistics of the noise terms $\xi_{1,2}$, which are given in terms of the mode functions of the quantum field, $\phi_{\bf k}$. These mode functions in turn depend on a specific realization of the background in which they evolve, through their self-energy, which acts as a $\phi_c$-dependent mass term.

A word on the precise structure of the perturbative expansion: we will solve the system $\{ \phi_c , \xi \}$ perturbatively in the number of quantum fields, $N_q$, and in the slow-roll parameter $\epsilon$. For example, the solution for $\phi_c$ at order $( \epsilon^0 , N_q = 0)$ corresponds to a constant background field with no stochastic corrections. A calculation of the quantum mode function can be done in this background, which is now at order $( \epsilon^0, N_q = 1)$. An equivalent counting is in powers of $\sqrt{\hbar}$, which counts loops in the Schwinger-Keldysh  formalism, i.e. $N_q = 1$ corresponds to $\mathcal{O}(\sqrt{\hbar})$. This was shown in \cite{Levasseur:2013ffa}. Given a quantum mode valid at order $\sqrt{\hbar}$, the variance of the noise can consistently be computed at order $\hbar$, and hence the PDF of IR modes can also be computed at order $\hbar$. In particular, the variance of IR modes (which encodes the spectral tilt), is valid at order $\hbar$. We will use this notation extensively in this paper.

\section{First Step of the Recursion: Stochastic Noise}
\label{sec:noise}

To make progress with the system (\ref{eq:phic}) and (\ref{modefunctioneom}) while maintaining the consistency of the solution, and in order to capture the fact that the quantum modes sit in a stochastic background, we use the recursive method from \cite{Levasseur:2013ffa}. At step 0 of the method, we start by approximating the background to zeroth order in slow-roll (i.e. pure de Sitter space). Step one of the method is to find the amplitude of the noise in this zeroth-order background.

More precisely, we want to calculate the amplitude of the classical noise arising from quantum fluctuations evolving in such a background and joining the coarse-grained theory. That is, we want to calculate $\xi_{1,2} = \xi^{(1)} _{1,2}$ to order $\{ \epsilon^0, \sqrt{\hbar} \}$ by solving equation (\ref{modefunctioneom}) and then use equation (\ref{themonstercorrelationmatrix}) to find the statistical properties of the noise terms.  Fortunately, it is possible to show (after simple algebra) that $\xi_2$ is always suppressed by some power of the slow-roll parameters, since it is proportional to at least one time derivative of the quantum field mode function $\phi_k$, and hence, at this order in perturbation theory, it is sufficient to calculate $\xi_1$ alone.

We start by making an explicit choice of the window function
\begin{equation}
\label{windowfunction}
W(k,t) = \theta(k - \gamma a H )\, ,
\end{equation}
where $\gamma\ll1$ parametrizes how long after their Hubble crossing modes can be considered classicalized, and so acts as an `ignorance parameter' (however, see \cite{Casini1999, Winitzki2000,  Liguori2004} for discussions of subtleties concerning this choice and explorations of different possibilities). 

Using this, and changing time variables to the number of $e$-folds (for reasons that will be discussed in sections~\ref{sec:step2} and~\ref{sec:step4}), the variance can be computed to the required order in slow-roll at this stage of the recursive method
\begin{equation}
\langle \xi^{(1)}_1 ({\bf x},N) \xi^{(1)}_1 ({\bf x}',N')\rangle = \frac{\gamma^3 H^5}{2 \pi^2} \frac{\sin (\gamma a H r)}{\gamma a H r} a^3 \left.| \phi_{\bf k}(N)|\right._{k={\gamma aH}}^2 \delta(N-N') ,
\end{equation}
where we have used the PDF (\ref{eq:masterPDF}), leaving the mode function unspecified in (\ref{theMmatrices}), to explicitly calculate the variance of the noise $\xi_1$, keeping only terms of order $\epsilon^0$ (where we define the first slow-roll by $\epsilon = - \dot{H}/H^2$). Here the modes $\phi_{\bf k}$ need to be evaluated at the time when they join the coarse-grained scales, which can be evaluated using the usual expression for mode functions in the $\mathcal{O}(\epsilon^0)$ de Sitter background,
\begin{equation}
\left.| \phi_{\bf k}(N)|\right._{k={\gamma aH}}^2 = \frac{H^2}{2 (\gamma a H)^3} ,
\end{equation}
and it follows that the mean and variance of the noise are given by
\begin{equation}
\label{stats}
\langle \xi_1^{(1)} ({\bf x},N)\rangle =0 \;\;\;,\;\;\; \langle \xi_1^{(1)} ({\bf x},N) \xi_1^{(1)} ({\bf x}',N')\rangle = \frac{H^4}{4 \pi^2} \frac{\sin (\gamma a H r)}{\gamma a H r} \delta(N-N') .
\end{equation}
The variance is constant and proportional to $\delta(N-N')$, and hence $\xi_1$ acts as white Gaussian noise\footnote{This will not remain true at higher orders in the recursive method: the noise will become colored due to interactions of the bath and the system, as was discussed in \cite{Starobinsky1994}.} with zero mean. The noise variance is local in time, and although it might appear to be nonlocal in space, the $\frac{\sin (\gamma a H r)}{\gamma a H r}$ factor in fact acts as a theta function at the coarse-graining radius, being one within the coarse-graining length, and zero outside. This ensures that the noise is only ($100\%$) correlated within each coarse-graining region, but is not correlated between different regions. Equivalently, this can be stated by saying the nonlocalities are only within the coarse-graining scale, and so the coarse-grained theory remains local.

\section{Step Two: Stochastically-Corrected Coarse-Grained Theory}
\label{sec:step2}
\subsection{Analytic solution}
\label{subsec:step2analytic}
In the previous subsection we assumed a classical nondynamical background and used it to calculate the noise $\xi_1 = \xi^{(1)} _1$ to order $(\epsilon^0, \hbar^{1/2} )$. Using this, we can calculate the corrected classical background $\phi^{(1)} _c$ at order $(\epsilon^1, \hbar)$. To do this, we solve equation (\ref{eq:phic}), which is a Langevin equation for $\phi^{(1)}_c$ at this order \footnote{Recall, as mentioned above equation (\ref{windowfunction}), that the noise terms $\dot{\xi}_1$ and $\xi_2$ are higher order in slow-roll.}:
\begin{equation}
\label{eq:timeSDE}
\dot{\phi} ^{(1)}_c\left({\bf x}, t \right) +\frac{ V_{,\Phi} \left(\phi_c^{(1)}\left({\bf x}, t \right) \right)}{3H_0} = \xi^{(1)} _q \left({\bf x}, t \right)\, .
\end{equation}
In the above, all quantities are valid to zeroth order in slow-roll. In particular, as should be explicit from the previous section, the variance of the noise term $\xi_1$ is valid to leading order in $\hbar$ but to zeroth order in slow-roll. Consistent perturbation theory then requires that the Hubble parameter appearing in (\ref{eq:timeSDE}) also be evaluated at zeroth order in slow-roll, and hence is simply constant.

Solving (\ref{eq:timeSDE}) gives a solution valid at $\mathcal{O}(\epsilon, \hbar)$. At this stage in the recursive method, we are considering quasi-de Sitter space rather than a nondynamical de Sitter spacetime. Gauge fixing therefore becomes necessary and we choose to work with gauge-invariant variables. In the stochastic formalism, this can be achieved by using the number of $e$-folds elapsed since the beginning of inflation, $N$, as the time variable. Recalling that we are working with the potential
\be
V(\Phi) = \frac{1}{2} m^2 \Phi^2,
\ee
we get
\be
\label{eq:1stvarphiSDE}
	\frac{\mathrm{d} \phi^{(1)}_c({\bf x}, N)}{\mathrm{d}N}= -\frac{ m^2 \phi^{(1)}_c({\bf x}, N)}{3H_0^2}+\frac{\xi_1^{(1)} ({\bf x}, N)}{H_0}  \, .
\ee
The advantage of using $N$ as the time variable is that linear order perturbations of the resulting stochastic process $\phi^{(1)}_c$ then coincide with the Mukhanov gauge-independent variable, as shown in \cite{Finelli:2008zg}. This is because in terms of the number of $e$-folds, Taylor expanding to linear order the full fields equations of motion yields the gauge-fixed perturbation equations. This will be discussed further in section~\ref{sec:step4}.

The solution to equation (\ref{eq:1stvarphiSDE}) can easily be written in terms of an integral equation:
\begin{equation}
	\label{eq:sol1of1realisation}
	\phi^{(1)}_c({\bf x},N) = \phi^{(1)}_c ({\bf x},0)\exp\left[-\frac{m^2}{3H_0^2}N \right]+\frac{H_0}{2\pi}\exp\left[- \frac{m^2}{3H^2 _0}N\right]\int e^{\left[\frac{m^2}{3 H^2_0}N \right]} \tilde \xi({\bf x},N) dN\, ,
\end{equation}
where we have made the rescaling $\xi_1^{(1)} = \frac{H_0^2}{2\pi}\tilde{\xi} $, and $\tilde\xi$ is therefore a regular Brownian motion with unit variance. From this, along with equation (\ref{stats}) for the statistics of the noise, we see that the incoming quantum modes leave the mean (i.e. the first-order moment) of the effective classical background unaffected and only modify higher moments of the background PDF.

Other derived stochastic quantities can also be calculated from the coarse-grained field $\phi_c^{(1)}$ at this order. For example, the slow-roll parameter $\epsilon$ will now have a stochastic piece (the same is true of the time-dependent Hubble parameter $H(t)$), which can be expressed as:
\bea
\label{eq:StochasticCorrectionToEpsilon}
	\epsilon\equiv-\frac{\dot H(t)}{H(t)^2}&=&\frac{\left(\dot\phi_c^{(1)}\right)^2}{2H^2(t)} - \frac{\xi_1\dot\phi_c^{(1)}}{2H^2(t)}= \left( \frac{\mathrm{d}\phi_c^{(1)}}{\mathrm{d}N}\right)^2\frac{H_0^2}{2H^2(t)}-\xi_1\frac{\mathrm{d}\phi_c^{(1)}}{\mathrm{d}N}\frac{H_0}{2H^2(t)}\, .
\eea
This can be rewritten as,
\be
\label{eq:epsilonstochastic}
	\epsilon=\epsilon^C+\epsilon^\xi=\frac{m^2}{3 H_0^2}-\frac{H_0}{2\pi}\frac{\tilde\xi}{\phi_c^{(1)}}\, ,
\ee
where we have separated the first slow-roll parameter into a classical piece $\epsilon^C$ and a stochastic piece $\epsilon^\xi$ . 

It is important to note that (\ref{eq:sol1of1realisation}) is not the most general expression to characterize solutions to (\ref{eq:1stvarphiSDE}), since each individual solution is only one realization the stochastic process $\phi_c^{(1)}$, as is made explicit by the presence of the Wiener process $\tilde\xi$ in the expression. Alternatively, we can solve for the probability density function $\rho(\phi_c^{(1)}({\bf x},N))$, which gives the probability of a field configuration over the whole length of inflation, using a Fokker-Planck equation.
Another, perhaps simpler, option is to solve (\ref{eq:1stvarphiSDE}) numerically, by solving many different realizations and from there inferring the shape of the underlying PDF using Bayes' theorem. This can be done by maximizing the likelihood on the $\mu$-$\sigma$ space (provided we assume the PDF is Gaussian) or, in the absence of Gaussianity, by finding the 68\% confidence levels. In the following, this is the strategy we will adopt.

\subsection{Numerical solution to the coarse-grained theory}

\noindent In terms of the rescaled noise $\tilde\xi$, the stochastic differential equation (SDE) to solve is:
\bea
	&\frac{d \phi_c^{(1)}}{dN}= -\frac{ m^2\phi_c^{(1)} }{3H_0^2}+\frac{H }{2 \pi}\tilde\xi ({\bf x}, N)  \,, \\
	&\langle \tilde\xi(N) \tilde\xi(N') \rangle=\delta(N-N') \, .\label{1stnoise}
\eea
In order to discretize the SDE, we need to discretize the time delta function in ($\ref{1stnoise}$):
\be
	\delta(N-N')=\left\{ \begin{array}{ll} 1/\delta N,&\quad\mathrm{if~}N\mathrm{~and~}N'\mathrm{~are~in~the~same~time~step~}\delta N\, ,\\
	0 ,& \quad\mathrm{otherwise.}\end{array} \right.
\ee
Here, $\delta N$ is the integration time step used in the numerical solver. This simple SDE can be solved using the Euler method to integrate:
\be
	\phi^{(1)}_{n+1}=\phi^{(1)}_{n}-\left[\frac{m^2\phi^{(1)}_n}{3 H_0^2}\right]\delta N+\frac{H_0}{2\pi} \tilde\xi_n\, ,
\ee
where the $\tilde\xi_n$ are independent random numbers drawn from a random normal distribution with standard deviation~$\sqrt{\delta N}$.

After simulating a large number of realizations of this coarse-grained background, the underlying PDF of the random variable $\phi_c^{(1)}$ can be reconstructed. This can be done by assuming an underlying Gaussian PDF and sampling the likelihood of the $\mu$-$\sigma$ space parametrizing the possible Gaussians to find the maximum likelihood. This means, at each time step, and for every plausible value of $\mu$ and $\sigma$, we apply Bayes theorem to find the probability that values from all realizations of $\phi_c^{(1)}$ are drawn from the Gaussian defined by a given choice of $\mu$ and $\sigma$.

\begin{figure*}[t]
\begin{center}
\centering
\begin{minipage}[t]{0.48\linewidth}
\centering
\includegraphics[trim = 80 -32 0 50 mm, width=1.02\textwidth]{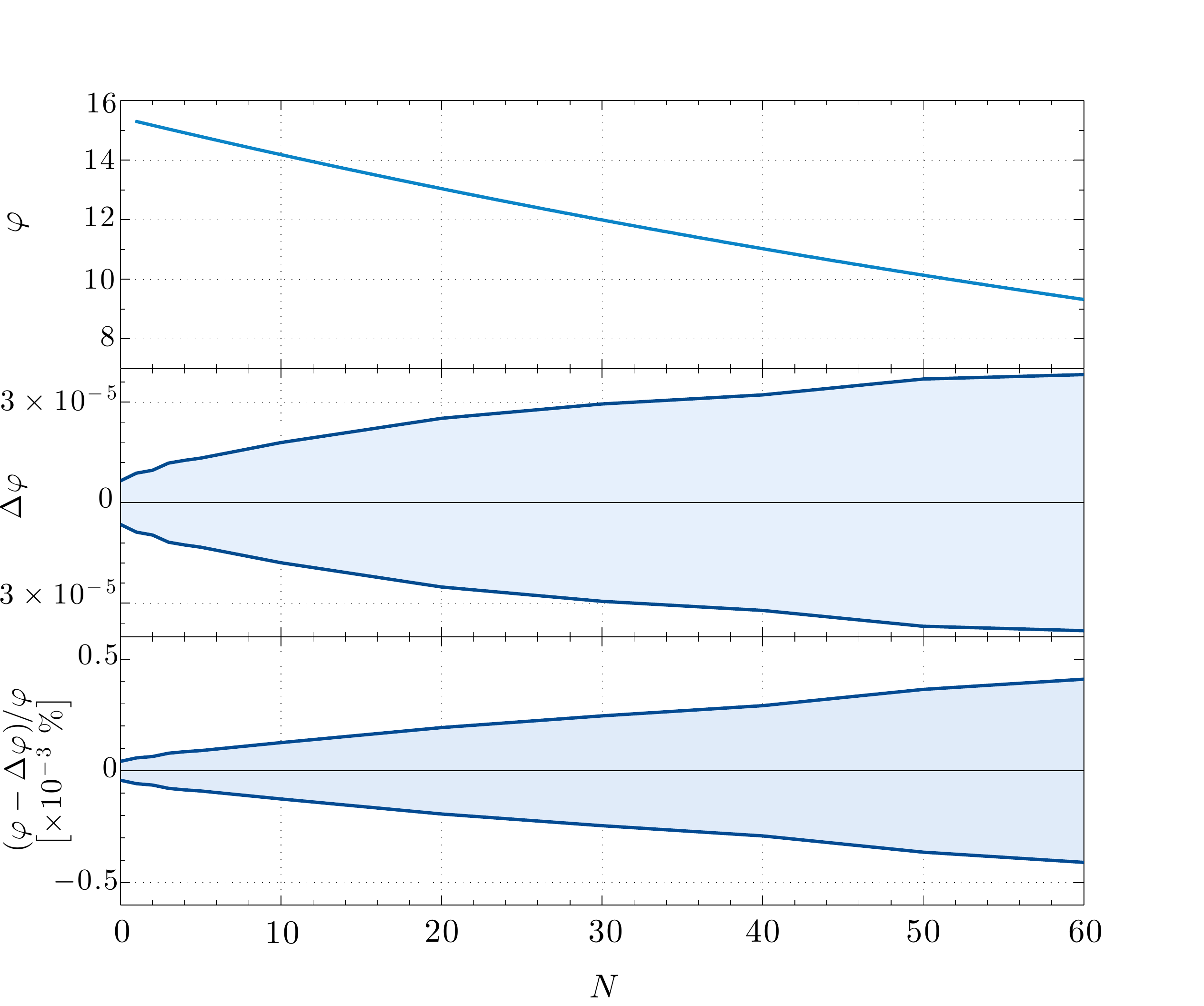}\\
\end{minipage}
\begin{minipage}[t]{0.48\linewidth}
\centering
\includegraphics[trim = 49 0 50 0 mm, clip, width=1.32\textwidth]{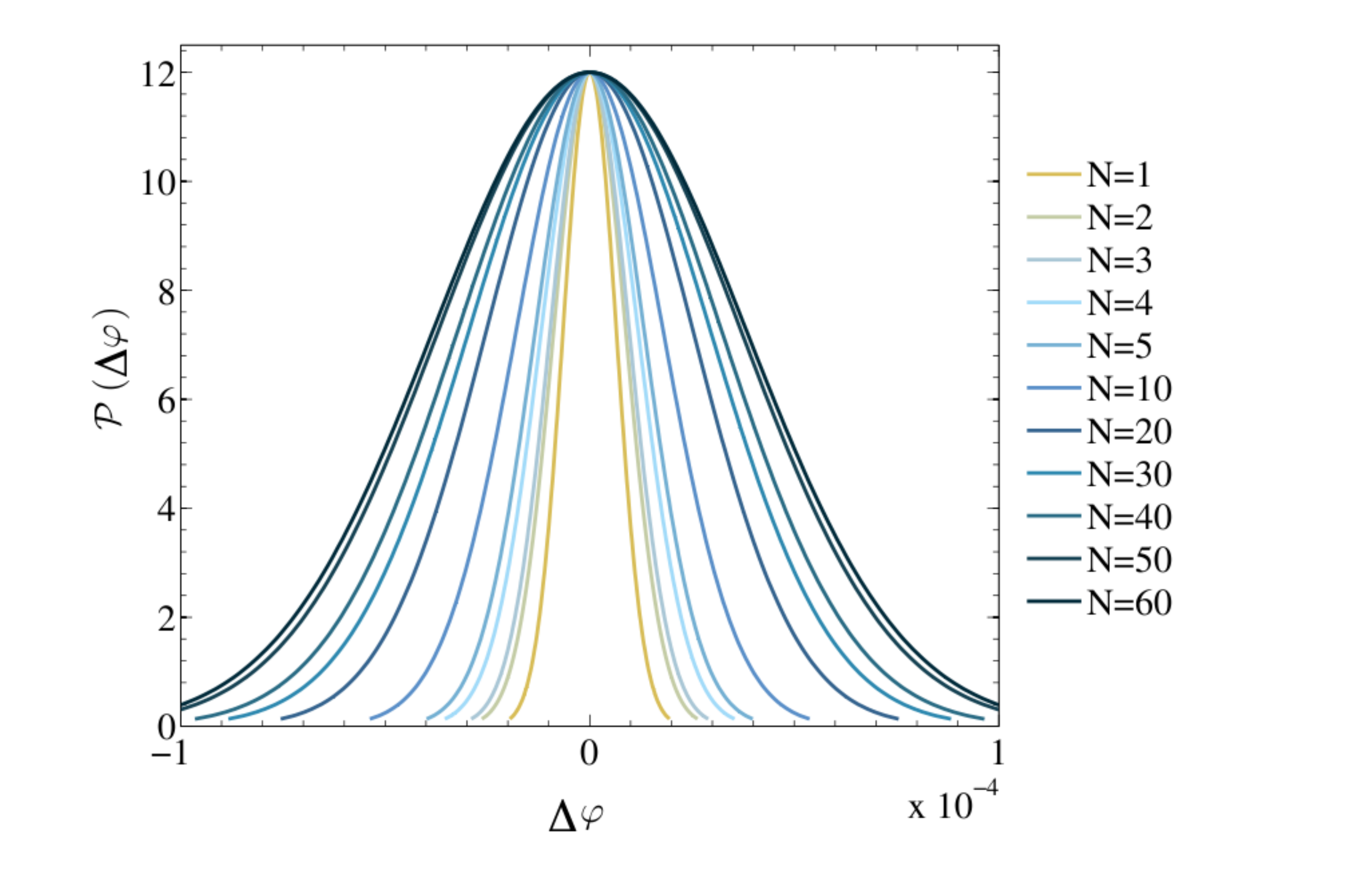}\\
\end{minipage}
\caption{\label{f:variations} 
{\it Left panel}: Top: mean trajectory of the coarse grained inflaton $\varphi$. Note that the large discrepancy of $\varphi(N=60)$ compared to the standard slow-roll value is simply due to the fact that at this level in the recursion, we solved the EoM for $\varphi$ keeping $H=H_0$, a constant. This means that by the end of inflation, we can expect corrections to our coarse-grained solution as large as $\epsilon_1N$. Middle: 1$\sigma$ deviation from the mean trajectory, i.e. standard deviation of maximum likelihood. Bottom: percentage error at 1$\sigma$. {\it Right panel}: The reconstructed (un-normalized) PDF of the random variable $\varphi(N)-\langle \varphi(N)\rangle$ for a few fixed $e$
-folds during inflation, and assuming inflation lasted 60 $e$-folds. These values were inferred from 250 realizations, with $m=6\times10^{-6}\Mp$, $N_{tot}=60$, and $\phi_c(0)=\sqrt{2(2N_{tot}-1)}$. All fields shown are in units of $\Mp$.
}

\end{center}
\end{figure*}

The resulting PDF for a few fixed $e$-folds over the course of inflation are shown for the $m^2 \phi_c^2$ potential in Figure~\ref{f:variations}. One can see that the variance of the long-wavelength field is initially zero, as one would expect from the fact that the PDF was initialized as a delta function in probability space. Its variance then grows as more modes exit the horizon and join the coarse-grained theory, which can be seen in the middle panel on the left-hand side (l.h.s.), as well as the plot on the r.h.s. The fractional variance of the field also grows during this time period, as shown in the bottom panel of the l.h.s.

The absolute variance does not grow indefinitely. As can be seen from both the l.h.s. and r.h.s. panels, the variance saturates during the last 10 $e$-folds of inflation, approaching a maximal value that can easily be estimated from (\ref{eq:1stvarphiSDE}):
\be
\label{step2variance}
	\sigma^2_{\phi_c^{(1)}}\equiv\left\langle ( \phi_c^{(1)})^2\right\rangle-\left\langle ( \phi_c^{(1)})\right\rangle^2~\rightarrow~\frac{3H_0^4}{8\pi^2m^2}\, .
\ee	
 Since the field is massive, this is as one should expect: quantum fluctuations becoming classical push the field fluctuations to roll up their potential, but the shape of the potential tends to make the field roll back down to its minimum. These two competing effects eventually reach an equilibrium point, which can be calculated in standard perturbative analysis to coincide with (\ref{step2variance}).
 
 As a final comment, note that, (\ref{step2variance}) being a constant, the power spectrum of curvature perturbations that we obtain at this stage in the recursive method is exactly flat. That is to say, the maximal equilibrium value that the field fluctuations reach is constant with time, which is consistent with a constant push from the incoming quantum modes. If the spectrum were tilted, this would correspond to kicks with time-dependent amplitude, and this would in turn modify the quasiequilibrium position for $\sigma^2_{\phi_c^{(1)}}$ , making it time dependent. This is what we will observe in the next level of recursion.

\begin{figure*}[t]
\begin{center}
\centering
\begin{minipage}[t]{0.48\linewidth}
\centering
\includegraphics[trim = 42 0 50 0 mm, clip, width=0.95\textwidth]{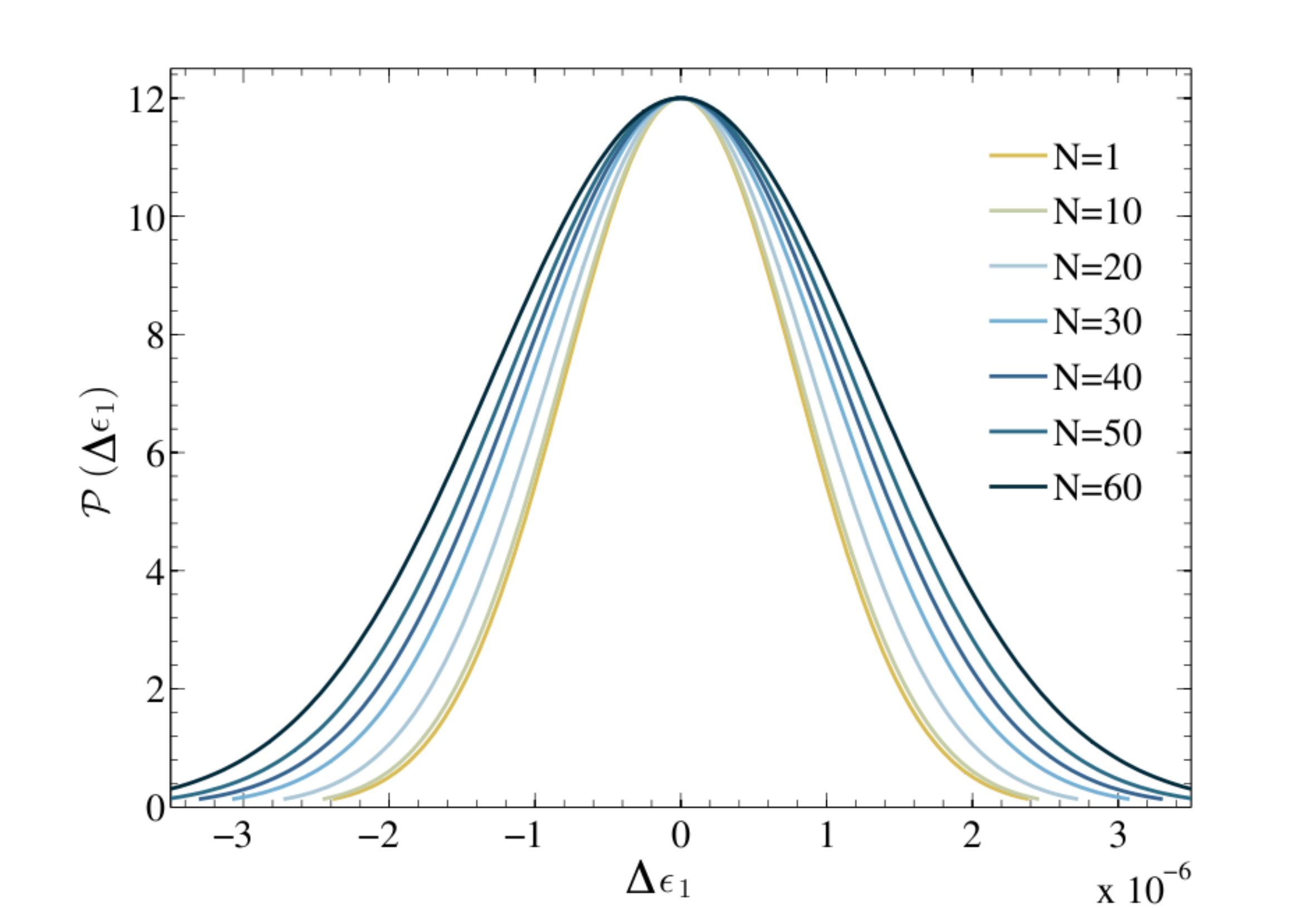}\\
\end{minipage}
\caption{\label{f:vareps} 
The reconstructed (un-normalized) probability density function of the slow-roll parameter $\epsilon_1$ (the subscript 1 denotes first slow-roll parameter) for a few fixed $e$-folds during the last 60 $e$-folds of inflation. These values were inferred from 250 realizations, with $m=6\times10^{-6}\Mp$, $N_{tot}=60$.
}

\end{center}
\end{figure*}

We can apply the same method to infer the underlying PDF of the slow-roll parameter $\epsilon$, displayed in Figure~\ref{f:vareps}, which exhibits a qualitative behavior similar to $\phi_c$. These graphs depict an interesting perspective: the super-Hubble classical theory we obtain is a `fuzzy' one, in the sense that the classical parameters have an inherent uncertainty stemming from the constant incoming quantum modes. Therefore, on the large scales of the coarse-grained theory (by which we mean on scales of many Hubble volumes), the value of $\epsilon$ varies from point to point with a standard deviation shown in Figure~\ref{f:vareps}.

Furthermore, even at a \emph{single point} the value of the classical parameters, such as $\epsilon$, are constantly fluctuating. In particular, this means that, when one averages over macroscopic timescales, there is a minimum possible value for the slow-roll parameter $\epsilon$. Indeed, even in the limit where $H_0$ (or equivalently $\phi_0$) starts out very large, in such a way that $\epsilon^C$ as defined in (\ref{eq:epsilonstochastic}) tends to 0, the contribution from $\epsilon^\xi$ will always remain finite. This is true in general for any model of inflation: the root-mean-square of $\epsilon^\xi$ always provides a minimal value of the first slow-roll parameter, regardless of how small it is engineered to be classically. It is worth stressing how the picture that we obtain differs from the standard one: the super-Hubble theory is now fundamentally probabilistic, and each realization of the quantum modes in the bath sees one of this ensemble of fluctuating field trajectories as a background.

\section{Step Three: Quantum Fields Evolving on a Stochastic Background}
\label{sec:step3}
\subsection{Analytic solution}

Now that a solution to (\ref{eq:phic}) valid to $\mathcal{O}\left(\epsilon, \hbar \right)$ has been found, we can go back to the bath (i.e. short-wavelength) sector of the theory and solve the linearized mode function of the quantum field to $\mathcal{O}\left(\epsilon, \hbar \right)$. This will allow us to find the noise variance to leading order in slow-roll. 

To do this, we treat the twice-corrected quantum modes, which we denote $\phi^{(2)} _q$, as perturbations about a fixed background field $\phi^{(1)} _c$.  It is important to note that this procedure requires a careful treatment of metric perturbations, the necessity of which was realized in \cite{Geshnizjani:2004tf}, where the authors considered the backreaction of cosmological perturbations (i.e. the effect of second-order perturbations on the background) in the presence of stochastic effects. However they did \emph{not} consider the backreaction of the shifted background on the noise itself, which is \emph{precisely} what we are interested in here. 

Before we progress further, we must first fix a gauge. The most natural choice is the same gauge as classical perturbations of $\phi_c$, that is, the spatially flat gauge. Following the treatment of \cite{Gordon:2000hv}, we fix the gauge to the spatially flat gauge, and find the equation of motion for the field perturbations:
\begin{equation}
\label{eompart3}
\ddot{\phi}^{(2)}_q + 3 H \dot{\phi}^{(2)} _q + \left[ -\frac{\nabla^2}{a^2} + m^2 - \frac{1}{\Mp^2 a^3} \frac{\mathrm{d}}{\mathrm{d}t} \left( \frac{a^3 \left(\dot{\phi}^{(1)}_{c}\right)^2}{H}\right)\right] \phi^{(2)} _q =0 \, ,
\end{equation}
It is important to note that $\phi_q ^{(2)}$ is \emph{a corrected version} of $\phi_q ^{(1)}$, as opposed to an additional contribution to the mode function $\phi_q$. Hence, we impose the Bunch-Davies initial condition on $\phi_q ^{(2)}$.

Solving the gauge constraints, transforming to Fourier space, and using the canonically normalized variable $v_k=a\phi_{\bf k}$, we find the equation of motion:
\begin{equation}
\label{eomstep3}
	v''_k-\left\{ k^2 - \frac{2}{\tau^2}-\frac{9\epsilon}{\tau^2}+\frac{V_{\Phi\Phi}(1+\epsilon)^2}{H^2(t)\tau^2}+4\frac{\dot H(t)}{H(t)}\frac{\ddot{\phi}^{(1)}_{c}}{\dot\phi^{(1)}_{c}}\frac{(1+\epsilon)^2}{H^2(t)\tau^2} -14\frac{\epsilon^2}{\tau^2} \right\}v_k=0\, ,
\end{equation}
where prime denotes a derivative with respect to $\tau$,  the conformal time defined by $a(t)\mathrm{d}\tau=\mathrm{d}t$. The solution to this, at first order in slow-roll, and after matching to Bunch-Davies initial conditions, is given by:
\begin{equation}
v_k = \frac{\sqrt{\pi}}{2} \sqrt{-\tau} H^{(1)} _{\nu} (- k \tau) \, ,
\end{equation}
\begin{equation}
\nu^2 = \frac{9}{4} + 9 \epsilon - \left(\frac{m}{H} \right)^2 (1+ 2\epsilon) \, ,
\end{equation}
where we have ignored the time dependence of $\nu$ coming from $\dot{\epsilon} \sim \dot{\xi}_1$ since it is suppressed relative to $\epsilon$. 

Now that we have an expression for the mode function evolution, we can obtain the expression for the noise variance, $\langle \xi_1, \xi_1 \rangle=\langle\xi_1^{(2)}\xi^{(2)}_1 \rangle$ valid to order $(\epsilon^1, \hbar^2)$. After some computation, see Appendix \ref{varcalc}, and doing a combined expansion in $(m/H)^2$ and $\epsilon$, the variance simplifies to (at first order in both $\epsilon$ and $(m/H)^2$)
\begin{equation}
\label{step3variance}
\langle \xi^{(2)}_1(N) \xi^{(2)}_1(N')\rangle = \frac{H^4(t)}{4 \pi^2} \left[ 1+\Delta \right] \delta(N-N') \, ,
\end{equation}
where $\Delta$ is defined as
\begin{equation}
\label{eq:Delta}
\Delta = \frac{2}{3} \left( \frac{m}{H}\right)^2 \left( - 2 + \gamma_E + \log 2 \gamma \right) - 3 \epsilon (-3 + 2 \gamma_E + 2 \log 2  \gamma)\, .
\end{equation}
Here, $\gamma_E$ denotes the Euler-Mascheroni constant. Note that the Hubble parameter $H$ appearing in the above two equations stands for the full $H$, i.e. stochastically and slow-roll corrected. Stochastic corrections to the variance of the noise are therefore included in $\epsilon$, as per equation (\ref{eq:StochasticCorrectionToEpsilon}), as well as in $H$. 

Also, note that in order to impose that the variance of the noise is independent of the choice of coarse-graining radius, that is, independent of $\gamma$, we must impose the hierarchy $\exp\left[-\frac{H^2}{m^2}\right]\ll\gamma\ll 1$ \cite{Starobinsky1994}. This is consistent with the expressions found in~\cite{Levasseur:2013tja}, and ensures that the effective theory we obtain through the coarse-graining process is a sensible and physical one.

\subsection{Numerical solution to the mode function equation}

To proceed with the numerical solution, we first recast equation (\ref{eompart3}) in a more useful  form. This equation can easily be rewritten in terms of the number of $e$-folds $N$:
\be
\label{modefunctionStep3}
	\frac{\mathrm{d}^2 v_k}{\mathrm{d}N^2}+\frac{\mathrm{d} v_k}{\mathrm{d}N}+\left\{ \frac{k^2e^{-2N}}{H^2(t)}(1+\epsilon)^2 - 2-9\epsilon+\frac{V_{\Phi\Phi}}{H^2(t)}(1+\epsilon)^2+4\frac{\dot H(t)}{H(t)}\frac{\ddot{\phi}^{(1)}_{c}}{\dot\phi^{(1)}_{c}}\frac{(1+\epsilon)^2}{H^2(t)} -14\epsilon^2\right\}v_k=0\, ,
\ee
where a dot refers to a derivative with respect to cosmic time $t$. Retaining only terms up to leading order in $\epsilon^C$, the classical piece of $\epsilon$, and up to $\mathcal{O}(\tilde\xi^2)$, we obtain the equation
\bea
	\frac{\mathrm{d}^2 v_k}{\mathrm{d}N^2}+\frac{\mathrm{d} v_k}{\mathrm{d}N}&+&\left\{ \frac{k^2e^{-2N}}{H^2(t)}(1+2\epsilon^C) - 2-9\epsilon^C_1+\frac{m^2}{H^2(t)}(1+2\epsilon^C_1)\right\}v_k\nonumber\\
&+&\frac{H_0}{\pi}\left\{9-2\frac{m^2}{ H^2(t)}-2\frac{k^2e^{-2N}}{ H^2(t)} \right\}\frac{\tilde\xi}{\phi^{(1)}_c}v_k+\frac{H_0^2}{\pi^2}\left\{ \frac{k^2 e^{-2N}}{H^2(t)}-14+\frac{m^2}{H^2(t)} \right\}\left(\frac{\tilde\xi}{\phi^{(1)}_c}\right)^2v_k=0\,.
\eea

To solve for the PDF of the stochastic linearized quantum mode function corresponding to (\ref{modefunctionStep3}) is very difficult, since (\ref{modefunctionStep3}) is now proportional to the square of the noise\footnote{In this case, an analytical solution for the PDF of the mode functions through a Fokker-Planck equation is not possible anymore.}. We therefore proceed numerically, using a modified version of the Runge-Kutta method for solving SDEs (which reduces to the improved Euler method in the absence of a stochastic term), as explained in Appendix B.

\begin{figure*}[t]
\begin{center}
\centering
\begin{minipage}[t]{0.48\linewidth}
\centering
\includegraphics[trim = 63 -32 0 50 mm, width=1.05\textwidth]{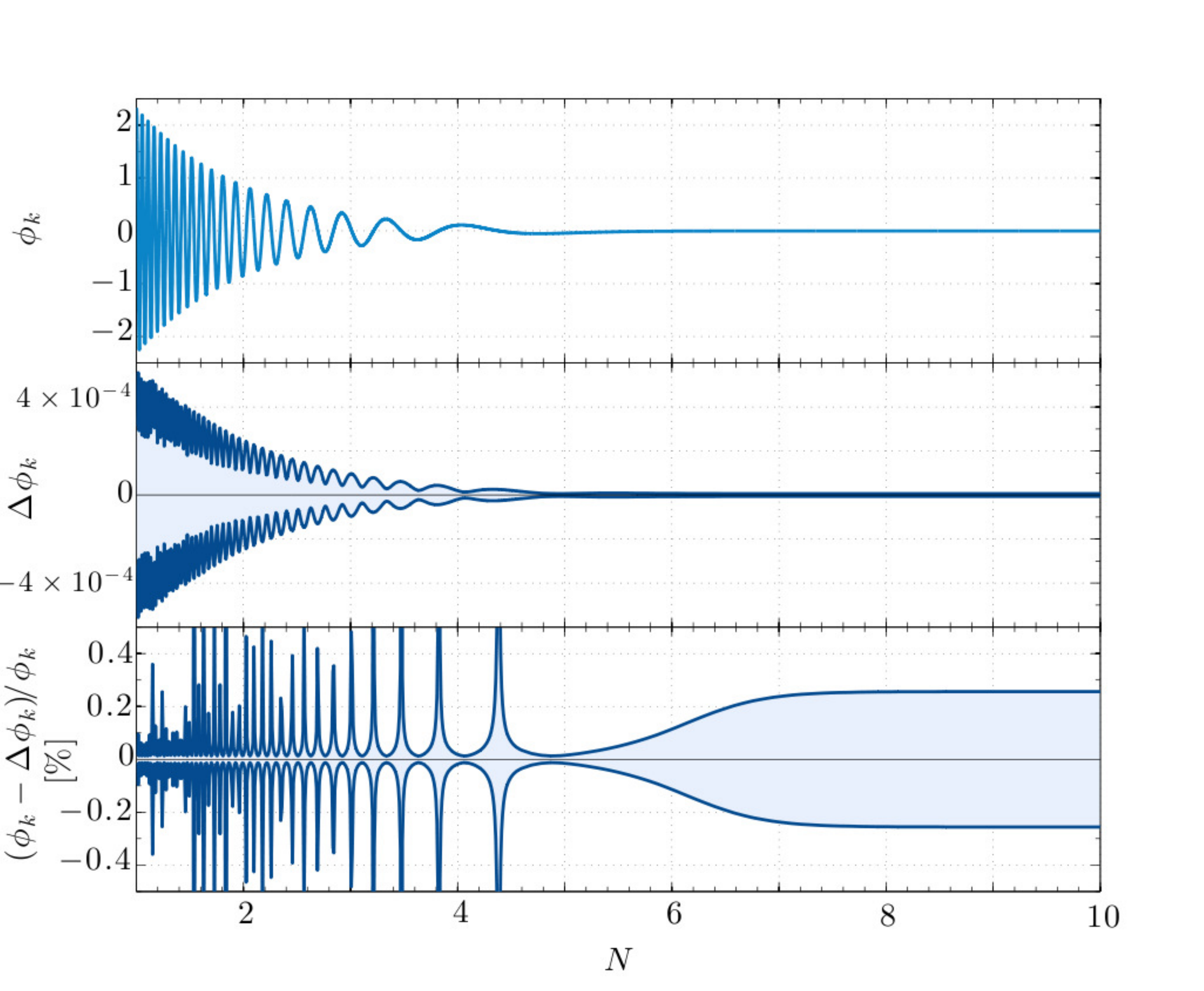}\\
\end{minipage}
\begin{minipage}[t]{0.48\linewidth}
\centering
\includegraphics[trim = 48 0 50 0 mm, clip, width=1.30\textwidth]{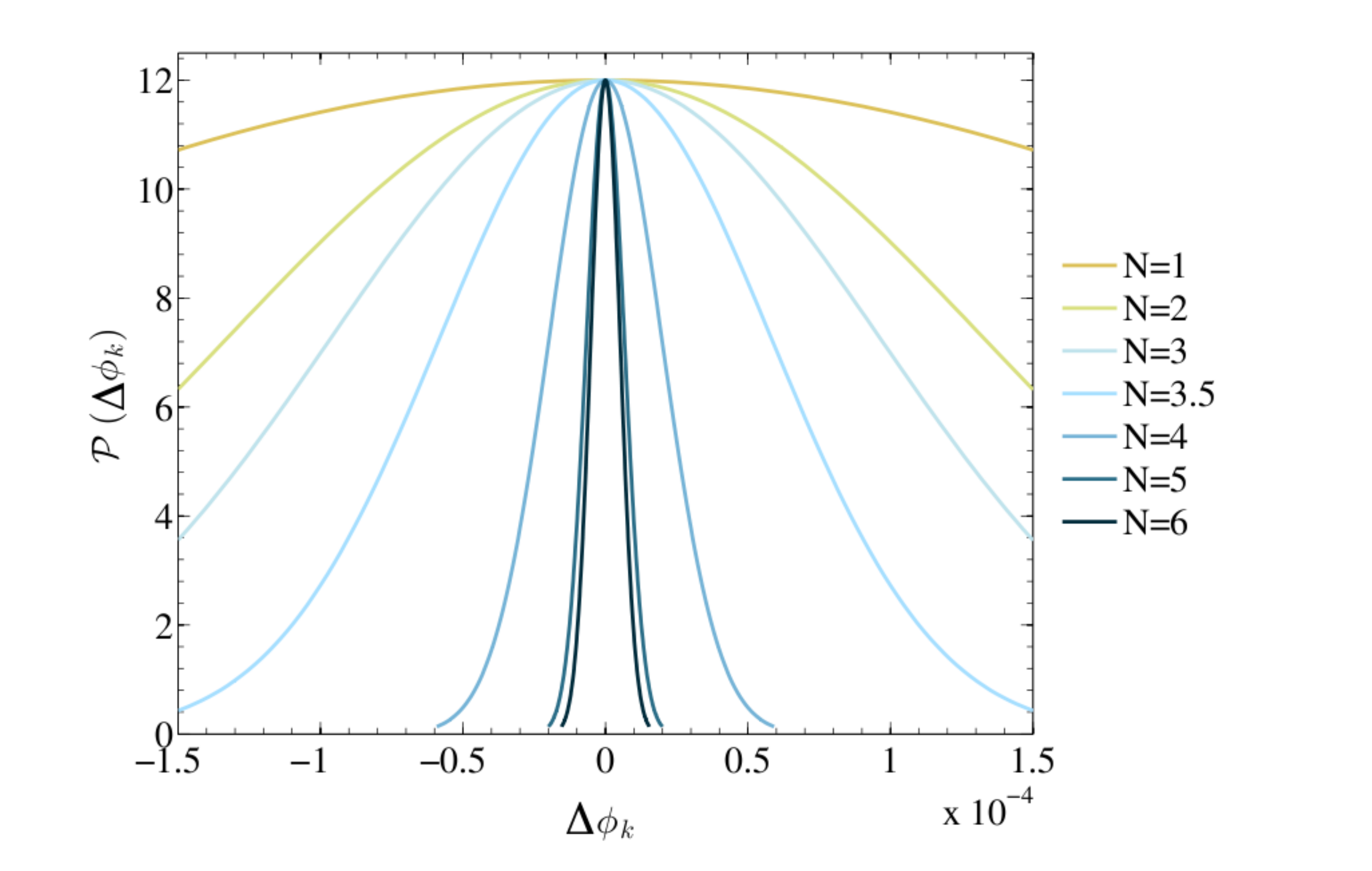}\\
\end{minipage}
\caption{\label{fig:modefunctionstep3} 
{\it Left panel}: Top: mean trajectory of the real part of the linearized mode function $\phi_k$ with $k=1.26\times10^{-2}$$\Mp$, freezing at $N\approx5.5$. Middle: 1-$\sigma$ deviation form the mean trajectory, i.e. standard deviation of maximum likelihood. Bottom: percentage error at 1$\sigma$. {\it Right}: The reconstructed (un-normalized) PDF of the random variable $\phi_k(N)-\langle \phi_k(N)\rangle$ for a few fixed $e$
-folds during the last 60 $e$-folds of inflation. These values were inferred from 250 realizations, with $m=6\times10^{-6}\Mp$, $N_{tot}=60$. Fields are shown in units of $\Mp$.
}

\end{center}
\end{figure*}

\begin{figure*}[t]
\begin{center}
\centering
\begin{minipage}[t]{0.60\linewidth}
\centering
\includegraphics[trim = 0 0 50 0 mm, clip, width=1.05\textwidth]{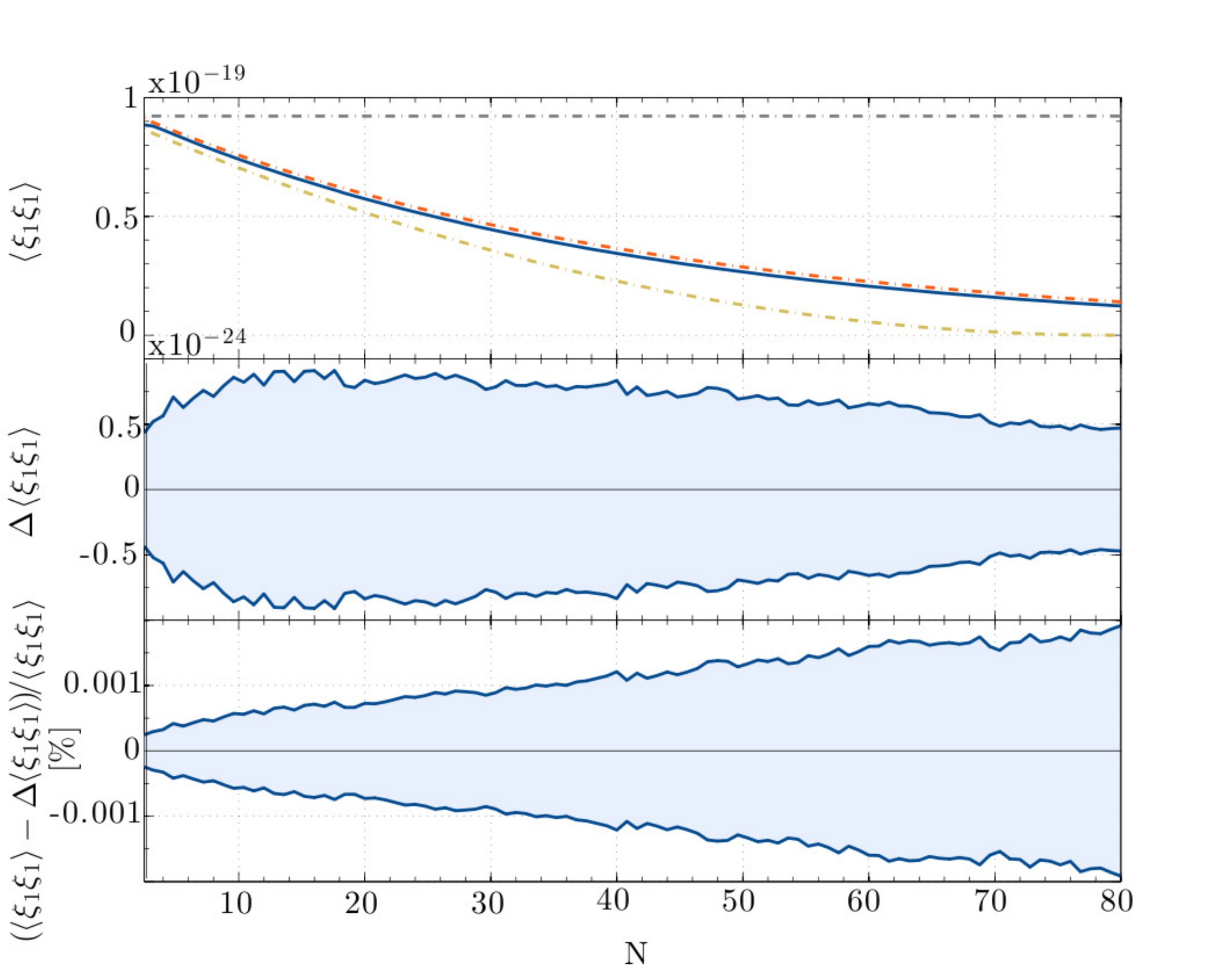}\\
\end{minipage}
\caption{\label{fig:errornoisePS} 
Power spectrum of the noise correlator at step 3 of the recursive method, $\langle\xi_1\xi_1 \rangle$. {\it Top panel}:  Average power spectrum, $\langle\xi_1\xi_1 \rangle$, in units of $\Mp^4$. The full (blue) line is the corrected power spectrum obtained numerically at step 3, while the red line is the analytic computation. The grey dashed line is the variance obtained at the previous step of the recursion, $H_0^4/(4\pi^2)$, and the yellow dashed line is $H^4(t)/(4\pi^2)$, i.e. the naive correction to the noise to account for the fact that inflation happens is quasi-de Sitter space by simply making the Hubble radius a time-dependent quantity. {\it Middle panel}: 1$\sigma$ deviation from the mean trajectory, i.e. standard deviation of maximum likelihood, again in units of $\Mp^4$. {\it Bottom panel}: Percentage error at 1-$\sigma$. These graphs were obtained by simulating 100 realizations, and, assuming a Gaussian underlying PDF, the mean and variances were inferred by maximizing the likelihood of the $\mu-\sigma$ space. Simulations were performed using $m=6\times10^{-6}\Mp$, $N_{tot}=80$.
}

\end{center}
\end{figure*}

In order to solve for \emph{each} realization of the mode function in a given realization of the background (solved for in step two of the recursive method, see section~\ref{subsec:step2analytic}), for every realization, the background and the mode function equation must be solved simultaneously (such that each realization of the mode function `sees' the background generated by the right Wiener process $\tilde \xi$).  The result for a fixed mode is displayed in Figure~\ref{fig:modefunctionstep3}.

Figure 3 highlights the generic behavior of a mode of the gauge-invariant Mukhanov variable $\phi_k$ inside and outside the Hubble horizon. Before horizon crossing, the mode $v_k$ has a constant norm. Therefore, $\phi_k$ plotted here has an amplitude that decays as $a^{-1}$. The reason why it appears to be oscillating widely in the top panel of Figure~\ref{fig:modefunctionstep3} is that only the real part to the mode function is shown, and the real and imaginary parts oscillate at identical speed with a $\pi/2$ phase shift. Around $N=5$, the mode plotted crosses the Hubble radius and freezes, and its amplitude remains constant from then on (real and imaginary parts independently). 

The absolute variance of the mode also decays as $a^{-1}$ while the mode is inside the horizon, as shown in the middle panel of the l.h.s., as well as the plot on the r.h.s. The fractional variance, displayed in the bottom panel of the l.h.s., diverges each time the real part of the mode crosses zero (which should not be interpreted as a physical effect), However,  after the mode has exited the horizon, both the field and the variance approach a constant (this can be seen in the r.h.s. and the middle panel of the l.h.s.), and the fractional variance converges to roughly 0.25\%. Note that, for our purpose in the present paper, we only apply the description of $\phi_k$ as a UV mode up until this mode joins the coarse-grained theory via the noise $\xi$, which occurs a few $e$-folds after horizon exit to ensure classicalization, around $N=7$ in Figure~\ref{fig:modefunctionstep3}.

Repeating this procedure for every $k$ mode exiting the coarse-graining radius during the last 60 $e$-folds of inflation in a given realization of the background, we obtain the corrected power spectrum of the stochastic noise. Figure \ref{fig:errornoisePS} shows the resulting average noise correlator (thick blue, top panel), when averaging over 100 realizations, and the 1-$\sigma$ error on this correlator on the middle (absolute) and bottom (fractional) panels. The red dot-dashed line in the top panel, representing the analytical calculation from equation (\ref{eq:Delta}), shows very good agreement between our numerical and analytical treatments. The top panel of the figure also shows, for comparison, the result at zeroth order in slow-roll which was obtained in section \ref{sec:noise} and used in section \ref{sec:step2}, as well as the naive slow-roll correction obtained by taking $H\rightarrow H(t)$ in the zeroth-order result, as one would obtain by following the procedure of, e.g.,~\cite{Finelli:2008zg} (yellow line).

\section{Step Four: Corrected Coarse-Grained Theory}
\label{sec:step4}
\subsection{Overview and numerical approach}
We have thus far completed two levels of recursion: 1) for our first `guess', we began with a nondynamical de Sitter background, then calculated the amplitude of the noise generated by quantum modes evolving on such a background in section \ref{sec:noise}; 2) using this noise (valid to leading order in $\hbar$ and zeroth order in slow-roll) as a source, we went back to the large scales and solved for the statistics of the coarse-grained classical inflaton, $\phi_c^{(1)}$, in section \ref{sec:step2}. Using this as a background (valid to first order in slow-roll and $\hbar$) for the short-scale physics, we then evaluated the corrected quantum modes in section \ref{sec:step3}. This then allowed us to find the variance of the noise arising from this bath, $\langle \xi_1\ \xi_1 \rangle=\langle\xi_1^{(2)}\xi^{(2)}_1 \rangle$, valid to $\mathcal{O}(\epsilon^1, \hbar^2)$.

Next, we shall use this noise to, once more, come back to the large-scale physics and source the coarse-grained theory. This will allow us to obtain a coarse-grained field $\phi_c^{(2)}$ valid to $\mathcal{O}(\epsilon^2,\, \hbar^2)$. That is, we must now solve:
\be
\label{eq:step4eom}
	\frac{\mathrm{d} \phi^{(2)}_c}{\mathrm{d}N} = - \frac{V_\Phi\left(\phi^{(2)}_c\right)}{3 H^2} + \frac{\xi^{(2)}_1}{H}\, ,
\ee
where now $H=m\phi^{(2)}_c/\sqrt{6}\Mp$ and $\xi^{(2)}_1$ is a random Gaussian variable sampled from a distribution with mean 0 and variance given by equation (\ref{step3variance}).

Recall that, although it should be thought of as a background when discussing the short-scale dynamics of the quantum mode functions inside the bath, the resulting $\phi_c$ is \emph{not} homogeneous, i.e. the stochastic contribution to $\phi_c$ is inherently inhomogeneous. Rather, the PDF for $\phi_c$ contains all the information about the classicalized field, including perturbations. This is an elegant way to encode a large amount of information; however, we are left with the problem of calculating the standard phenomenological parameters of inflationary cosmology, such as the spectral tilt. 

Numerically, however, solving this equation is quite easy. Using a method analogous to what was done in section \ref{sec:step2}, we solve for each realization of the coarse-grained theory using a realization of the noise output that was used toward the construction of Figure~\ref{fig:errornoisePS}. After constructing 100 realizations of $\phi_c$, we then use Bayes theorem to infer the two first moments of the underlying PDF (assuming Gaussianity), i.e. its mean and variance as functions of time. 

\begin{figure*}[t]
\begin{center}
\centering
\begin{minipage}[t]{0.60\linewidth}
\centering
\includegraphics[trim = 0 0 50 0 mm, clip, width=1.05\textwidth]{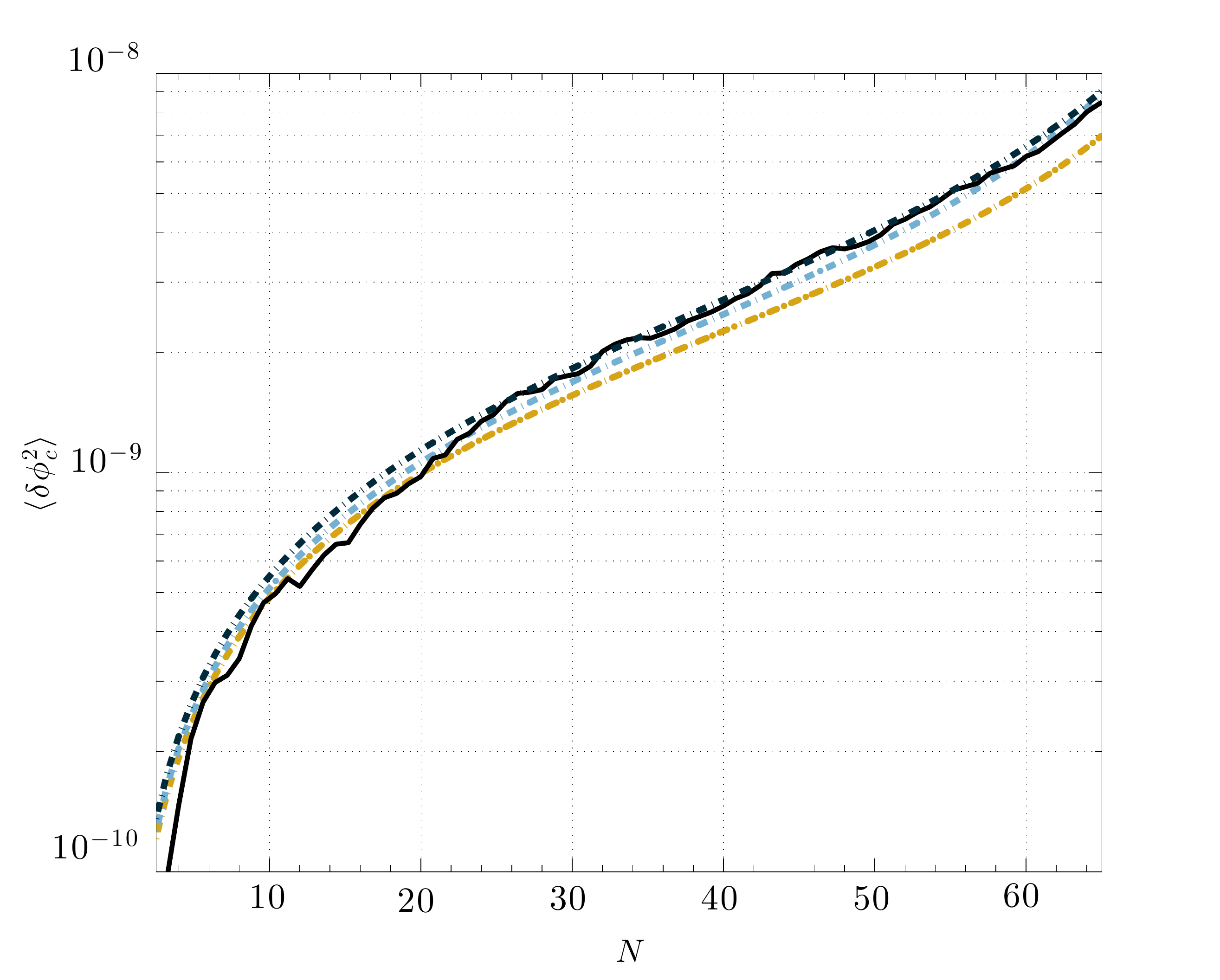}\\
\end{minipage}
\caption{\label{fig:classPS} 
Total integrated power in, or variance of, the classical long-wavelength fluctuations, $\langle \delta \phi_c ^2 \rangle$, in units of $\Mp^2$. The full (black) line is the numerical result, while the dashed light- and dark-blue dashed lines correspond to different analytic approximations to the integral in equation (\ref{eq:classfieldfluc_th}), see footnote 8 for details.  The yellow dashed line corresponds to a noise variance of $H^4(t)/(4\pi^2)$, i.e. the naive approximation obtained by slow-roll corrected $H(t)$ to account for the fact that the bath evolves in a slow-rolling background~\cite{Finelli:2008zg}.
 This graph was obtained by simulating 100 realizations, and, assuming a Gaussian underlying PDF, the mean and variances were inferred by maximizing the likelihood of the $\mu-\sigma$ space. Simulations were performed using $m=6\times10^{-6}\Mp$, $N_{tot}=80$.
}

\end{center}
\end{figure*}

As one should expect with a choice of parameters excluding eternal inflation, this additional step in the recursive method does not give significant corrections to the mean trajectory of the inflaton. Its variance, however, is the quantity capturing the integrated power of the classicalized field fluctuations, and is of great interest to us. This is the quantity presented in Figure \ref{fig:classPS}, where the numerical result is the black solid line.

\subsection{Inflaton fluctuations beyond leading order}

\subsubsection{Perturbation equations for the random variable $\phi_c$}

As alluded to in section \ref{sec:step2}, the analysis of fluctuations in stochastic inflation can be done quite simply by using the number of $e$-folds $N$ as the time variable. This fact can also be seen by considering the $\delta N$ formalism. As an example, let us consider single field inflation, following \cite{Sugiyama:2012tj}. The background equation of motion is given by (where $\rho$ is the energy density):
\begin{equation}
3 H^2 \Mp ^2  = \rho ,
\end{equation}
\begin{equation}
H \partial_N (H \partial_N \phi) + 3 H^2 \partial_N \phi + \partial_ \phi V(\phi) =0 .
\end{equation}
The power of the $\delta N$ formalism comes from realizing that, in the absence of entropy perturbations, the above equation applies \emph{nonperturbatively}. This leads to a statement of the `separate universe approach' to perturbations,
\begin{equation}
\phi (N) = \phi_0 (N, \phi_{init}, (\partial_N \phi)_{init} ) .
\end{equation}
This equation states that the nonperturbative dynamics in one region of spacetime are captured by solving for the background $\phi_0$, but given a set of perturbed initial conditions $\{ \phi_{init}, (\partial_N \phi)_{init} \}$. In fact, the full $\delta N$ formalism is much more powerful than this, as it is easily generalizable to a gradient expansion. 

This same formalism can (and has, see \cite{Fujita:2013cna, Fujita:2014tja}) be applied to stochastic inflation, since the noise evolves independently in different Hubble patches, and hence does not spoil the `separate universe approach'.\footnote{To be more precise: fluctuation and dissipation terms that arise at 3rd order are \emph{non-local}, meaning this approach would need to be modified. However, these non-localities are at most at the coarse-graining radius, and therefore do not spill into neighboring Hubble patches. Furthermore, here we are considering a free scalar field, and are neglecting the coupling between tensor and scalar perturbations.} In fact, this locality was shown to be a necessary condition for stochastic inflation in \cite{Tolley:2008na}. The applicability of the $\delta N$ formalism could in principle be shown more rigorously by expanding the action for the stochastic inflation, which was derived in \cite{Levasseur:2013ffa}. 

This approach can be applied here as follows. The equation of motion for the classical coarse-grained field $\phi^{(2)}_c$ is given by equation (\ref{eq:step4eom}). We can split the field $\phi^{(2)}_c$ into a homogeneous mode, which is just the expectation value $\langle \phi^{(2)}_c \rangle$, and an inhomogeneous piece $\delta \phi_c$ containing all the \emph{classical} fluctuations.\footnote{We will not write a subscript $^{(2)}$ on the inhomogeneous piece of $\phi_c^{(2)}$ for the sake of simplifying the notation.} We therefore expand (\ref{eq:step4eom}) around $\langle \phi_c^{(2)} \rangle$ using:
\be
  \phi^{(2)}_c = \langle \phi^{(2)}_c \rangle + \delta\phi_c\, ,
\ee
to find the equation for the first-order fluctuations~\cite{Finelli:2008zg}:
\begin{equation}
	\label{eq:phifluct}
	\frac{\mathrm{d} \delta\phi_c}{\mathrm{d}N}+2\Mp\left(\frac{H_{,\langle \phi^{(2)}_c\rangle}}{H}\right)_{,\langle \phi_c^{(2)}\rangle}\delta \phi_c=\frac{\xi^{(2)}_1}{H}\, ,
\end{equation}
where, as in equation (\ref{eq:step4eom}), $H$ is defined as including \emph{both} the slow-roll and stochastic corrections. The above equation can be solved to give the PDF for $\delta \phi_c$, from which we would like to extract information about the power spectrum. 

Alternatively, the variance of $\delta \phi_c$ can be calculated from (\ref{eq:phifluct}) by multiplying both sides by $\delta \phi_c$ and averaging, without having to solve for the full PDF of the classical field,
\be
	\label{eq:phifluct2}
	\frac{\mathrm{d} \langle \delta\phi_c^2\rangle }{\mathrm{d}N}+4\Mp\left(\frac{H_{,\langle \phi^{(2)}_c\rangle}}{H}\right)_{,\langle \phi^{(2)}_c\rangle}\langle \delta \phi_c^2\rangle =\frac{S}{4\pi^2H},\qquad \mathrm{with} \qquad S=\left(\left.H\right|_{\phi_c^{(1)}}\right)^3 (1+\Delta)\, ,
\ee
where, apart for the occurrences of $H$ in $S$, all other powers of $H$ are evaluated at $\langle\delta\phi^{(2)}_c\rangle$. To arrive at this equation, we have used the relation $\langle\delta\phi_c \xi_1\rangle=(\left.H\right|_{\phi_c^{(1)}})^3(1+\Delta)/(8\pi^2)$, which can be deduced by expanding $\xi_1$ and $\delta\phi_c$ in terms of their Fourier modes, and enforcing continuity of the (amplitude of the) full field $\Phi$ across the horizon $k=\gamma a H$. We emphasize that the cumbersome notation is necessary: the slow-roll correction to the variance of the noise was calculated with respect to the background $\phi_c^{(1)}$, where as the occurrences of $H$ in equation (\ref{eq:step4eom}) are defined with respect to $\phi_c^{(2)}$.

This equation is easily solved in terms of the homogeneous solution for $\langle \phi_c \rangle$,
\be
\label{eq:classfieldfluc_th}
	\langle\delta\phi_c^2\rangle=-\frac{H^2_{,\langle \phi^{(2)}_c\rangle}}{8\pi^2\left(\left.H\right|_{\langle \phi^{(2)}_c \rangle}\right)^2\Mp^2}\int\frac{\left.H\right|_{\langle \phi^{(2)}_c \rangle}}{H^2_{,\langle \phi^{(2)}_c\rangle}}S \mathrm{d}N \approx -\frac{H^2_{,\langle \phi^{(2)}_c\rangle}}{8\pi^2\left(\left.H\right|_{\langle \phi^{(2)}_c \rangle}\right)^2\Mp^2}\int\frac{\left(\left.H\right|_{\langle \phi_c \rangle}\right)^5}{H^3_{,\langle \phi_c\rangle}}(1+\Delta)\mathrm{d}\langle\phi_c\rangle\, .
\ee
In the last step, we have assumed that the time evolution of $\left.H\right|_{\langle \phi^{(2)}_c \rangle}$ and $\left.H\right|_{ \phi^{(1)}_c }$ are the same. To be precise, they differ by terms that are higher in slow-roll than the precision to which $\left.H\right|_{ \phi^{(1)}_c }$ is defined.

Substituting the form of $\left.H\right|_{\langle \phi^{(2)}_c \rangle}$ in the above equation yields the following solution for the total power in the fluctuations of the coarse-grained inflaton as a function of time:\footnote{If one had instead solved the integral exactly, i.e. \emph{kept} the occurrences of $\left.H\right|_{ \phi^{(1)}_c }$ in (\ref{eq:classfieldfluc_th}), the integral to solve would have had the form  $\sim\int_0^N \langle\phi_c^{(2)}(N')\rangle \phi_0^3\exp\left[-\frac{m^2}{H^2}N'\right]dN'$. Solving this integral, (\ref{eq:classfieldfluc_th}) would become $-\frac{i \phi_0^6 \left(\Gamma\left[\frac{3}{2}, -\frac{3}{2}\right] - 
      \Gamma\left[\frac{3}{2}, -\frac{3}{2} + \frac{6 N}{\phi_0^2}\right]\right)}{3 \sqrt{6} \e^{\frac{3}{2}}}
 \frac{m^4 \phi_0^3}{6^2}\frac{1}{4\pi^2}\frac{1}{\left.H\right|_{\langle \phi_c^{(2)}\rangle}^2}$. The theoretical prediction from this result for $\langle\delta\phi_c\rangle$ is shown in Figure~\ref{fig:classPS} (dark blue dashed line), and is negligibly close to the result from the expression in (\ref{stochasticstd_th}).}
\be
\label{stochasticstd_th}
	\langle\delta\phi_c^2\rangle=\frac{H_0^6-H^6}{8\pi^2m^2H^2}(1+\Delta)\rightarrow \frac{H_0^6}{8\pi^2m^2H^2}(1+\Delta)\, ,
\ee
where the right arrow denotes the asymptotic value approached towards the end of inflation (as the field approaches its minimum). 

The first equality in the above equation (before the limit is taken) is the analytic result of the recursive formalism, and is shown in Figure 4 (dashed dark-blue line). The pale-blue dashed line shows the first equality in equation (\ref{eq:classfieldfluc_th}), before any approximation on $\left.H\right|_{\langle \phi^{(1)}_c \rangle}$ is done. The good agreement between this, the final analytical result, and the variance of the classical field obtained numerically (full black line) supports the validity of our approach. For comparison, we have also plotted (dashed yellow line) the obtained total integrated power in classical fluctuations obtained in \cite{Finelli:2008zg, Finelli:2010sh}, by using the slow-roll corrected $H(t)$ to account for the fact that the bath evolves in a slow-rolling background (by enforcing it by hand). In contrast, the recursive method we apply here self-consistently accounts for this correction, in a natural way.

\subsubsection{Recovering the power spectrum of scalar fluctuations}

Clearly, when calculating the power spectrum of scalar perturbations, the often-used procedure of quantizing the fluctuations deep inside the horizon and then evaluating at horizon crossing cannot be applied here, as the quantum fluctuations have been replaced by a classical noise which, at every given time, is only nonzero at the coarse-graining scale. The power spectrum of perturbations can instead be calculated by noting that the variance of fluctuations is the integral of the power spectrum from an IR cutoff to the Hubble horizon, or more precisely, the classicalization radius, which corresponds to the coarse-graining scale. That is,
\begin{equation}
\label{integral}
\langle \delta \phi_c ^2\rangle = \displaystyle \int _{l} ^{\gamma a H} \mathcal{P}_{\delta \phi_c}(k) \; \mathrm{d} \log k \, ,
\end{equation}
which is a standard textbook result (see, for example,~\cite{textbook}).  The power spectrum in the above expression can be written in terms of mode functions as
\begin{equation}
	\mathcal{P}_{\delta \phi_c}(k) =\frac{k^3}{2 \pi^2} |\delta \tilde{\phi}_k|^2 \, ,
\end{equation}
where the tilde is to denote that $\delta \tilde{\phi}_k$ is not a Fourier mode of $\delta \phi_c$, but rather the mode function that one would find in the standard procedure of quantizing perturbations and computing the power spectrum.

Another approach consists in using the following trick: we can parametrize a generic power spectrum in terms of a general spectral index $n_s$, and explicitly compute the integral on the r.h.s. of equation (\ref{integral}). Then, by solving equation (\ref{eq:phifluct2}) for $\langle \delta \phi_c ^2\rangle$, we can deduce the value of $n_s$. More explicitly, in standard perturbation theory, a generic power spectrum of field fluctuations far outside the Hubble radius can be written as (see for example \cite{Martin:2003bt, Finelli2003}, where we keep only the terms that depend on $k$):
\be
	 \mathcal{P}_{\delta \phi_c}(k)=\mathcal{A}_s(t) k^{3} \left[1 + \frac{\left(n_s-1\right)}{2} \log \left(\frac{k}{k^*}\right)\right]^2\, ,
\ee
where $\mathcal{A}_s(t)$ is a time-dependent amplitude and $n_s$ is the spectral index.
Therefore, integrating over all super-Hubble modes as in (\ref{integral}), we obtain
\be
\label{eq:chap5Finellisspectrum}
	\langle \delta \phi_c ^2\rangle = \frac{1}{4\pi^2} \mathcal{A}_s(t) \frac{2}{3(n_s-1)}\left. \left[1 + \frac{\left(n_s-1\right)}{2} \log \left(\frac{k}{ k^*}\right) \right]^3 \right|^{\gamma aH}_{l}\, ,
\ee
where a standard calculation would take $\gamma =1$. Performing the renormalisation of this quantity through adiabatic subtraction as done in~\cite{Finelli2003}, we can obtain a value for $\langle \delta \varphi^2 \rangle_{\mathrm{REN}}$ that is independent of the IR cutoff. From there, evaluating this result after a sufficiently long period of inflation (in order for the one-point correlator $\langle \delta \varphi ^2\rangle$ to saturate to its maximal asymptotic value and for the memory of initial conditions to disappear), the terms in square brackets in equation~(\ref{eq:chap5Finellisspectrum}) (plus counter terms) simplify to $\left\{\left. \left[1 + \left(n_s-1\right) \log (\frac{k}{k^*}) \right]\right|^{\gamma aH}_{l}+\, c.t.\right\}\,\rightarrow - 1$. Therefore, given $\langle \delta \phi_c ^2\rangle_{\mathrm{REN}} $,  one can solve for the spectral index:
\be 
\label
{eq:magicformula}
	n_s -1=- \frac{1}{6 \pi^2}\frac{ \mathcal{A}_s(t)}{\langle \delta \phi_c ^2\rangle_{\mathrm{REN}} }\, .
\ee

\subsubsection{Computation of the time-dependent amplitude}
In the specific case of $m^2\Phi^2$ inflation, the standard theory of cosmological perturbations gives the following result for the time dependence of mode functions far outside of the Hubble horizon:
\be
	\delta\tilde\phi_k= \frac{1}{a^\frac{3}{2}}\left(\frac{\pi (1+\epsilon)}{4 H(t)}\right)^{\frac{1}{2}}\left( \frac{H(t_k)}{H(t)}\right)^2 \mathcal{H}^{(1)}_{3/2} \left((1+\epsilon) \frac{k}{aH}\right)\, , \qquad \mathrm{with} \qquad H(t_k)=H_0\sqrt{1+2\frac{\dot H_0}{H_0^2}\log\left( \frac{(1+ \epsilon_0) k}{H_0 \nu_0}\right) }\, ,
\ee
where $\mathcal{H}^{(1)}$ is the Hankel function of the first kind, and $\epsilon_0 = - \dot{H}/H_0 ^2$.
At small $-k \tau$ or $(1+\epsilon) \frac{k}{aH}$, i.e. outside the Hubble radius, the mode functions can be asymptotically approximated by:
\be
	\delta\tilde\phi_k\approx \frac{1}{(aH)^{3/2}}\frac{1}{2^{1/2}}\frac{1}{(1+\epsilon)}\frac{H^2_0}{H(t)}\left( 1-2\epsilon_0
	 \log\left(\frac{(1+\epsilon_0)k}{H_0\nu_0}\right)\right)\left(\frac{k}{aH}\right)^{-3/2}\, .
\ee
This allows us to calculate the r.h.s. of equation (\ref{integral}), which gives the total amount of power in the inflaton fluctuations:
\bea
	\displaystyle \int _{l} ^{\gamma a H} \mathcal{P}_{\delta \tilde{\phi}_c}(k)  \,\mathrm{d}  \log k	&=&  \frac{1}{4\pi^2}\frac{1}{(1+\epsilon)^2}\frac{H_0^4}{H^2(t)}\frac{(-1)}{6\epsilon_0} \left[\left( 1-2\epsilon_0  \log\left(a\frac{(1+\epsilon_0)H\nu}{(1+\epsilon)H_0\nu_0}\right)\right)^3-\left( 1-2\epsilon_0  \log\left(\frac{(1+\epsilon_0)l }{H_0\nu_0}\right)\right)^3\right]\, ,\nonumber \\
	&\rightarrow&  -\frac{1}{4\pi^2}\frac{1}{(1+\epsilon)^2}\left(\frac{H_0^4}{H^2(t)}\right)\frac{1}{3 (-2\epsilon_0)}\, .
	\label{calculatingA}
\eea
The arrow in the last line denotes the value at which the correlator saturates towards the end of inflation. From this, we can deduce an explicit form for the time-dependent amplitude of the power spectrum:
\begin{equation}
\label{amp}
\mathcal{A}_s(t) = \frac{1}{(1+\epsilon)^2}\left(\frac{H_0^4}{H^2(t)}\right) .
\end{equation}

\subsubsection{The spectral index}
\label{spectralindex}
The final result for  the stochastically-corrected spectral tilt $n_s$ can now be obtained from equations (\ref{stochasticstd_th}) and (\ref{eq:magicformula}), in combination with the above result (\ref{amp}) for the amplitude, to arrive at
\be \label{tiltcorrected}
	n_s-1\approx -4\epsilon_0 (1-2\epsilon)(1-\Delta) +\mathcal{O}\left(\left(\epsilon_1^2\right)_{\mathrm{std~pert}}\right) \, .
\ee
Here, $\mathcal{O}\left(\left(\epsilon_1^2\right)_{\mathrm{std~pert}}\right)$ represents additional second-order terms which appear in the standard slow-roll calculation, but do not bear any stochastic contributions. From this expression, one can see that $(n_s - 1)$ is manifestly negative, corresponding to a red tilt of scalar perturbations, as one would expect for this simple inflationary model.  Also, the specific value of the tilt to leading order in slow-roll matches that obtained from standard methods of calculations for $m^2\phi^2$ inflation.

 The corrections to the standard slow-roll result appear at second order in $\epsilon$. In particular, the slow-roll correction to the noise is captured by $\Delta$, which is easily checked to be positive definite. Hence we find a positive second-order correction to $n_s$.  This may come in addition to other second-order slow-roll corrections, as denoted by the term $\mathcal{O}\left(\left(\epsilon_1^2\right)_{\mathrm{std~pert}}\right)$, which come from the fact that we have neglected higher-order terms in (\ref{calculatingA}) when calculating $\mathcal{A}_s(t)$. Retaining those, it is in principle possible to recover standard second-order results (see, for example, \cite{Stewart:1993bc,Casadio:2005xv}), plus stochastic corrections. 

The additional effect of stochastic corrections to $n_s$ could have been anticipated from Figure~\ref{fig:classPS}, where the recursive calculations done in this paper are compared to the $\langle \delta \phi_c ^2\rangle$ one would get from simply taking $H\rightarrow H(t)$ in the variance of $\xi$. The recursive calculation is positively shifted with respect to the naive slow-roll correction, indicating that stochastic effects push the spectral index towards scale invariance, i.e., closer to zero. This is a general result for stochastic corrections in single field models: recursive corrections become more important in the late stages of inflation, and generically increase the variance of the field fluctuations. This is a second-order `blue' (i.e. positive) contribution to the spectral tilt.

Equation (\ref{tiltcorrected}) is the main result of this section. While the above correction to standard slow-roll inflation occurs only at second order in slow-roll, this serves as a proof-of-principle that stochastic corrections can indeed affect inflationary observables. As we will discuss in section \ref{multifield}, these corrections can be large in multifield models.

\section{ Inflation with Extra Heavy Fields}
\label{multifield}

The corrections due to stochastic effects can be more dramatic in multifield inflation. In particular, a direct coupling of the inflaton to extra heavy fields will give a stochastic correction to the mass of the inflaton, which enters the spectral index at first order in slow-roll. However, one should be careful not to choose the masses to be so large that their dynamics can simply be integrated out, i.e. we want to look at the case $m_\phi\ll m_\chi\lsim H$. This is the case in hybrid inflation, as has been studied in \cite{Levasseur:2013tja}, and can easily be generalized to $N$ fields.

Consider $m^2 \Phi^2$ inflation in the presence of a heavy field $\Psi$, which couples directly to the inflaton through a potential $V(\Phi,\Psi)$. For simplicity, let's take the potential 
\begin{equation}
V(\Phi,\Psi) = \frac{1}{2} g^2 \Phi^2 \Psi^2\, ,
\end{equation}
where $g$ is a dimensionless coupling constant. The classical dynamics is determined by the equations
\begin{eqnarray}
3 H^2 \frac{\mathrm{d} \Phi}{\mathrm{d}N}& =& - m^2 _{\Phi} \phi - g^2  \Phi \Psi^2\, ; \\
3 H^2 \frac{\mathrm{d} \Psi}{\mathrm{d}N}& =& - m^2 _{\Psi} \psi - g^2  \Phi^2 \Psi \, .
\end{eqnarray}
Let us consider a hierarchy of VEVs, that is $\langle \Phi \rangle \gg \langle \Psi \rangle$, which physically corresponds to a slowly rolling field $\Phi$ and the field $\Psi$ oscillating about its minimum. As we assume $\Psi$ is a relatively heavy field, the oscillations of $\Psi$ are suppressed by the large mass $m_\Psi$, and this in turn will induce corrections to spectrum of $\Phi$ that are suppressed by powers of $m_\Psi$. Similar scenarios have been discussed in many works, for example \cite{Chen:2009we, Chen:2009zp, Achucarro:2012sm, Achucarro:2012yr}.

However, the picture in stochastic inflation is quite different: for sufficiently large (though less than the Hubble energy scale) $m_\Psi$, the dynamics of the coarse-grained part of $\Psi$, $\chi$, can easily be dominated by stochastic effects. In this case, and since the field $\Psi$ has vanishing VEV, the value of the field is well characterized by its standard deviation:
\begin{equation}
\chi(N) \sim \sqrt{\langle \chi^2 \rangle} = \sigma_{\chi}.
\end{equation}
The equation of motion for $\phi_c$ can now be rewritten
\begin{equation}
3 H^2 \frac{\mathrm{d} \phi_c}{\mathrm{d}N} = - m^2 _{\Phi} \phi_c \left(  1 +\frac{g^2 \sigma_\chi^2}{m^2 _{\Phi}}\right) +3H\xi_1= - \tilde{m}^2 _\Phi \phi_c +3H\xi_1\, ,
\end{equation}
where we have defined the stochastic-corrected mass $\tilde{m}^2 _{\Phi} = m^2 _{\Phi}  +g^2 \sigma_\chi^2$. The recursive formalism then allows for a consistent and precise calculation of the statistics of $\phi_c$, as per the single field case.

This can be easily generalized to the case with a set of $N$ heavy fields, with variances denoted by $\sigma_i$,
\begin{equation}
\label{renormalizedmass}
\tilde{m}^2 _{\Phi} = m^2 _{\Phi}  + \sum_i g_{i}^2 \sigma_i ^2 ,
\end{equation}
and the spectral index of inflaton perturbations can be read off from the single field case,
\begin{equation}
n_s - 1 = - \frac{4 \tilde{m}^2 _{\Phi}}{3 H^2} (1 + \delta) \approx  -  \frac{4 {m}^2 _{\Phi}}{3 H^2} -  \sum_i \frac{4 g_{i}^2 \sigma_i ^2}{3 H^2}  ,
\end{equation}
where $\delta$ is a recursive correction, analogous to the single field case, that enters at second order in slow-roll, and the $\approx$ denotes that we have truncated to first order in slow-roll. We thus find a non-negligible correction to the inflaton spectrum in the presence of direct couplings to heavy fields, and interestingly, the stochastic correction in this case is red. 

Note that, here, we were careful to choose the number of fields and their masses in such a way that the inflaton itself always dominates the dynamics of the Hubble constant, even after the massive fields have developed a stochastic VEV. That is, writing the effective potential for the inflaton as $V=V_0+\frac{1}{2}\tilde m_{\Phi}^2 \Phi^2$, $V_0$ remains negligible throughout the analysis.

If, on the contrary, we had chosen the initial potential to have a large VEV, $V_0\gg m_{\Phi}^2 \Phi^2$ throughout inflation, the VEV that the massive fields develop due to stochastic effects would still renormalize the mass of the inflaton in a fashion identical to (\ref{renormalizedmass}). However, as in the standard analysis, the tilt we would obtain would be blue, and the stochastic corrections would make it bluer at linear order in slow-roll.

More subtle scenarios can also be examined, in particular two-field inflation with a turning trajectory \cite{Gao:2013ota}, and more general multifield inflation models~\cite{Gordon:2000hv}. We leave the analysis of such setups to future work.

\section{Tensor Fluctuations}

The analysis of scalar perturbations in the stochastic formalism can be straightforwardly extended to include tensor perturbations, as is most easily seen from the functional derivation~\cite{InPrep}. However, the phenomenology can be easily worked out without such detailed knowledge, by noting that tensor modes evolve as massless scalar fields. The analysis is nearly identical to the scalar case already studied. Tensor perturbations are defined as
\be
	ds^2=a^2(\tau)[d^2\tau^2-(\delta_{ij}+{\bf h} _{ij})dx^{i}dx^j]\, .
\ee
The second order action for dimensionless tensor perturbations is given by
\begin{equation}
	S=\frac{\Mp}{8}\int d\tau d^3x a^2\left[ \left({\bf h}'_{ij}\right)^2-\left(\nabla {\bf h}_{ij}\right)^2\right]\, .
\end{equation}
Assuming no anisotropic stress, the gauge-invariant Einstein equation for the tensor mode is given by
\be
	\ddot{\bf h}_{ij}+3H\dot{\bf h}_{ij}-\frac{\nabla^2}{a^2}{\bf h}_{ij}=0 .
\ee
Decomposing the tensor perturbations into eigenmodes of the Laplacian with ${\bf h}_{ij}={\bf h}^\lambda(t)e^\lambda_{ij}(x)$, where $\lambda~\varepsilon~\left\{ \times, +\right\}$ are the two polarization states of the fluctuations, and $\nabla^2 e^\lambda_{ij}=-k^2e^\lambda_{ij}$, it becomes easy to split the full field ${\bf h}_{ij}$ into a coarse-grained part, $h_{ij}$, and a bath part, $w_{ij}$,
\ba
	{\bf h}_{ij}&=&h_{ij}+w_{ij}\,, \\
	w_{ij}&=&\int \frac{d^3{\bf k}}{(2\pi)^3}\theta\left(\frac{k}{\gamma aH}-1\right)\sum_{\lambda=1,2} \hat{w}^\lambda_{{\bf k}}(\tau)e^\lambda_{ij}({\bf k})e^{i{\bf k\cdot x}}\nonumber \,.
\ea
Far outside the Hubble radius, the IR modes (which are part of the coarse-grained theory) are frozen out, and their equation of motion can therefore be approximated by
\begin{equation}
\label{thetensorstochasticqen}
\frac{\mathrm{d} h_{ij}}{\mathrm{d}N} = \frac{\xi_{h_{ij}}}{H},
\end{equation}
where $\xi_{h_{ij}}$ is the tensor noise. Following the scalar analysis, the variance of the noise at first order in slow-roll is given in terms of the tensor mode functions $w_{{\bf k}, \lambda}$ by
\begin{equation}
\label{tensorvariance}
\langle \xi_{h_{ij}}(N) \xi_{h_{ij}}(N')\rangle = \frac{(\gamma a)^3 H^5}{2 \pi^2} \sum_{\lambda=1,2} \left|w^\lambda_{{\bf k}}\right|^2 _{k=\gamma a H} (1- \epsilon) \delta(N-N') \, , 
\end{equation}
where we have used the polarization tensors identities $e^\lambda_{ij}(-{\bf k})=\left(e^\lambda_{ij}({\bf k})\right)^*$ and $e^\lambda_{ij}({\bf k},\lambda)\left(e^\mu_{ij}({\bf k})\right)^*=\delta^{\lambda\mu}$. 

From this, we can readily calculate the amplitude of the slow-roll corrected noise. To do so, we need to compute the mode functions $w^\lambda_{{\bf k}}$ evolving on the slow-roll and stochastically-corrected background calculated in section \ref{sec:step2}, $\phi_c^{(1)}$.  Defining the canonically quantized variable $u^\lambda_{{\bf k}} = \frac{a}{2} \Mp w^\lambda_{{\bf k}}$, we can write
\begin{equation}
\left(u^\lambda_{k}\right) '' + \left(k^2 - \frac{a''}{a}\right)u^\lambda_{{\bf k} } = 0 .
\end{equation}
Using $\frac{a''}{a} = (aH)^2 (2 - \epsilon)=\frac{2+3\epsilon}{\tau^2}$ and following the scalar case, this leads to the variance at first order in slow-roll,
\begin{equation}
\label{tensorvarianceSR}
\langle \xi_{h_{ij}}(N) \xi_{h_{ij}}(N')\rangle = \frac{2\left.H\right|_{\phi_c^{(1)}}^4}{\Mp ^2 \pi^2} [ 1 + \epsilon(1 - 2 \gamma_E - 2 \log 2 \gamma) ] \delta(N-N') , 
\end{equation}
where the additional factor of 2 comes from summing over possible polarization states.

Since the homogeneous mode of tensor perturbations is  zero, $\langle h_{ij} \rangle=0$, we can straightforwardly obtain an equation for the total amount of power in classical tensor fluctuations by a procedure analogous to what allowed us to go from (\ref{eq:phifluct}) to (\ref{eq:phifluct2}),
\be
	\frac{d \langle h_{ij}^2\rangle}{dN}=\frac{2}{\pi^2 \Mp ^2} \frac{\left.H\right|_{\phi^{(1)}_c}^3}{\left.H\right|_{\phi^{(2)}_c}} (1+\Delta)\, .
\ee
The solution to this equation is
\be
	 \langle h_{ij}^2\rangle=\frac{\sqrt{6}}{\pi^2}\frac{2H_0^3}{m \Mp^2}\int_0^N\frac{\exp\left[-\frac{m^2}{H_0^2}N' \right]}{\phi^{(2)}(N')}dN'=\frac{1}{\Mp^2}\frac{\sqrt{6}H_0^4}{\pi^2m^2}\frac{\sqrt{\pi}\left[  erfi\left(\sqrt{\frac{3}{2}}\right)-erfi\left(\frac{\phi_c^{(2)}(N)}{\phi^{(2)}_c(0)}\sqrt{\frac{3}{2}}\right)\right]}{e^{3/2}}\, ,
\ee
which can be expanded, to leading order in slow-roll,
\be
	\frac{\sqrt{6\pi}}{e^{3/2}}\left[  erfi\left(\sqrt{\frac{3}{2}}\right)-erfi\left(\sqrt{\frac{3}{2}}\frac{\phi_c^{(2)}(N)}{\phi^{(2)}_c(0)}\right)\right]\approx 2\left( 1- \left(\frac{\phi_c^{(2)}}{\phi_0}\right)^3\right) .
\ee
Taking the limit towards the end of inflation, we find
\be
 \langle h_{ij}^2\rangle\rightarrow 2\frac{H_0^4}{\pi^2 m^2 \Mp^2}\, .
\ee

On the other hand,  a general power spectrum of tensor fluctuations can be parametrized by
\be
	k^3 \mathcal{P}_h(k)=\mathcal{A}_T(t) k^{3} \left[1 + \frac{n_T}{2} \log \left(\frac{k}{\gamma aH}\right)\right]^2\, ,
\ee
with $\mathcal{A}_T(t) =8 \mathcal{A}_s(t)$ (as can be shown from a straightforward calculation, including the two possible polarizations of the tensor modes \footnote{Note that $\mathcal{A}_s$ refers to the amplitude of field fluctuations, which is related to the amplitude of curvature perturbations by $\zeta^2=\frac{1}{2\epsilon}\delta\phi^2$.}).
Therefore, we find that the tensor spectral index is:
\be
	n_T = -2\epsilon +\mathcal{O}(\epsilon^2)\,,
\ee
which is precisely the standard slow-roll result, see for example \cite{Stewart:1993bc}. The stochastic corrections to this can be calculated in a similar manner to the scalar case, and we leave this to further work. Note that, as in the scalar case, the stochastic correction will enter as a blue contribution to $n_T$, and hence pushes the spectral index towards scale invariance.\footnote{Note that our method can also capture the backreaction of tensor modes on scalar modes (which occurs at second order in standard perturbation theory) through the appearance of higher-order noise terms in (\ref{thetensorstochasticqen}), in a similar fashion to the additional noise terms which were shown to appear at the 1-loop level in \cite{Levasseur:2013ffa}. However, as shown in, e.g., \cite{Acquaviva:2002ud}, the induced second order perturbations have the form $\epsilon\left(\mathcal{R}^{(1)}\right)^2$, and so such an effect only appears at 3rd order in slow-roll. To the level of accuracy relevant for the present calculation, this effect is negligible.} As the precision of CMB B-mode experiments increases, these corrections (and, in general, precise phenomenology) will become an increasingly important consideration.

\section{Conclusion}
In this work we have put forward a detailed example of how the recursive formalism of stochastic inflation \cite{Levasseur:2013ffa,Levasseur:2013tja} can be applied to models of single field inflation. A key difference between our analysis here and the recursive prescription of the original papers is our treatment at step three of the recursive method. Indeed, when computing the dynamics of the bath's quantum modes on a stochastically corrected background, we did not impose that the modes evolve on a representative realization of the background (constructed by replacing all occurrences of the coarse-grained fields by their average in the bath propagator). Rather, we kept the background seen by these quantum modes purely stochastic. In \cite{Levasseur:2013tja}, this approach was sufficient to capture the leading contribution of stochastic effects to observables, since the dominant effect arose from a spectator field, while the inflaton was well approximated as deterministic. However, this statement is not true in general, and in the present analysis we went beyond this approximation in a fiducial example of a single field model.
 
 The picture we put forward here is that, during inflation, modes inside the Hubble horizon evolve on a background that is `fuzzy' in field space on scales~$\lambda< H_0/(2\pi)$. This is because, within each realization, the background seen by those modes is getting kicked at every moment in time, in a random direction with a rms of that size. This approach allows one to capture nontrivial stochastic corrections, even when the average of the coarse-grained field remains identical to that of the classical standard approach, or when no even powers of the coarse-grained fields\footnote{Recall that Gaussianity of the field implies that the expectation value of an odd power of fields is zero.} appear in the time-dependent mass of the quantum modes.

Throughout the present paper, we have studied, both analytically and numerically, the slow-roll correction to the growth of quantum fluctuations in $m^2 \phi^2 $ inflation away from the regime of eternal inflation, and found a nontrivial stochastic correction to their variance which could not have been deduced by simply correcting the Hubble constant to first order in slow-roll. We then used to this to compute the corrected long-wavelength modes, which includes both the slow-roll and stochastic effects. 

We separated the long-wavelength physics into a homogeneous component (i.e. the background) and an inhomogeneous component (i.e. primordial classical perturbations), from which we could compute the spectral tilt of classical scalar perturbations. We found a stochastic correction to the spectral tilt in $m^2 \phi^2$ inflation, which enters $n_s$ at second order in slow-roll, as a blue (i.e. positive) contribution. While this is a small effect in the case of a single field inflation, the same formalism can be straightforwardly applied to other models of inflation, and the machinery used to compute the corresponding corrections. One example is the case of inflation with direct couplings to heavy fields, wherein stochastic effects induce a shift in the effective mass of the inflaton, leading to a red contribution to the spectral tilt.

Extending the current analysis to the regime of eternal inflation would be an interesting followup to this work, however, it is not immediately straightforward because of a number of issues. The first and main reason for this is the need for a `reading rule' for the SDE, which can be explained as follows: as the eternal inflation regime is entered, the local value of $H$ and of the classical coarse-grained field $\phi_c$ become dominated by their stochastic contribution (that is to say, the time evolution of the coarse-graining radius itself becomes a stochastic process). To obtain the incoming noise at each time $N$, which is proportional to the local Hubble radius, one needs to know the amplitude of all the modes that previously crossed the Hubble radius\footnote{Note that this does not make the process non-Markovian, since the precise history of how $H(t)$ acquired that value is irrelevant, but to calculate $H(t)$ at a fixed time the prior history of a realization becomes essential.}. This can be summarized by saying that, in the eternal inflation regime, the noise is {\it multiplicative} (and cannot be considered additive), i.e., the importance of the noise term in the SDE depends on the stochastic process $\phi_c$ itself. 

This affects the previous analysis in the sense that it renders the SDE (\ref{eq:phic}) or (\ref{eq:1stvarphiSDE}) undefined without a so-called reading rule. That is, when defining a SDE as a limit of discretized equations as
\bea
	&\phi_c(N+\Delta N)-\phi_c(N)=-\left.\frac{V_{,\Phi}}{3H^2}\right|_{\phi_c(t)_\alpha} \Delta N+\left.\frac{H}{2\pi}\right|_{\phi_c(t)_\alpha}\int_N^{N+\Delta N}dN'\hat\xi(N')\, ,&\nonumber \\
	&\langle \hat\xi (N) \hat\xi(N') \rangle=\delta(N-N')\, ,&
\eea
where the weighted average $\phi_c(N)_\alpha$ is defined by:
\be
	\phi_c(N)_\alpha=(1-\alpha)\phi_c(N)+ \alpha\phi_c(N+\Delta N)\, ,
\ee
one must specify a value for $\alpha$ between 0 and 1. Failing to do so when the noise coefficient depends on $\phi_c$ explicitly makes the SDE undefined, since in general different reading rules (i.e. different choices of $\alpha$) give rise to different stochastic processes $\phi_c$ \cite{Gardiner1984}. Typically, $\alpha=0$ (referred to as an Ito process) corresponds to discrete systems, whereas $\alpha=1/2$ (referred to as a Stratonovich processes) corresponds to continuous physical systems; however, cases with exotic $\alpha$ values have also been found \cite{Volpe2010}. Determining the correct value of $\alpha$ for eternal inflation falls beyond the scope of this work, and we plan to return to this issue in a follow-up paper.  

In the last section of this paper, we finished by extending our analysis to include tensor perturbations, and again found a correction only at second order in slow-roll. Our analysis was simple, although a rigorous path integral derivation of this approach is still in progress.  We leave this, and the analysis of general multifield models, to future work. We conclude on one final remark: The formalism of stochastic inflation is an effective field theory, where the UV modes are integrated out, and an IR theory is obtained with extra operators (i.e. noise) capturing the effect of UV physics. However, in this case, the UV and IR sectors of the theory remain coupled, and the `cutoff' (the horizon) that separates these two sectors is time dependent. It follows that one must proceed with caution when solving this system.

\bigskip
\begin{center}
{\bf Acknowledgements}
\end{center}

The authors would like to thank Robert Brandenberger and  Anne Davis for enlightening discussion throughout the realization of this project and comments on different stages of the manuscript, as well as Vincent Vennin for useful discussion in the early stages of this work, and Elisa Gouvea Ferreira for comments on the initial draft. EM would like to thank \'Ecole de Physique Les Houches and the organizers of the Summer School 2013 for inspiring the inception of this project. LPL would like to thank McGill University and the Stanford ITP for their hospitality while part of this work was completed. EM would also like to thank the Stanford ITP for hosting him while a portion of this work was completed. The research of LPL and EM is supported in part by NSERC. The research of LPL is in addition supported by an M.T. Meyer scholarship, a Kyth\'e Waldram Travel Award, and a Rouse Ball Studentship in mathematics.

\renewcommand{\theequation}{A \arabic{equation}}
\renewcommand{\thefigure}{A \arabic{figure}}
\setcounter{equation}{0}  
\setcounter{figure}{0}
\setcounter{subsubsection}{0}
\appendix
\section{{\scshape Variance of the slow-roll corrected noise}} 
\normalsize

\label{varcalc}
The variance of the noise is given in terms of the mode function by
\begin{equation}
\label{appendixvariance}
\langle \xi(N) \xi(N')\rangle = \frac{\gamma^3 H^5}{2 \pi^2} a^3 |\phi_k|^2 _{k=\gamma a H} (1- \epsilon) \delta(N-N') , 
\end{equation}
where $H$ is the full stochastic and slow-roll correct Hubble constant, hence the corrected variance is itself stochastic. Now we get $|\phi_k|^2$
\begin{equation}
|\phi_k|^2 = |\frac{v_k}{a}|^2 ,
\end{equation}
where $v_k$ is the solution to the mode function equation with Bunch-Davies initial conditions.  The mode function is given by
\begin{equation}
v_k = \frac{\sqrt{\pi}}{2} \sqrt{- \tau} H ^{(1)} _{\nu} (- k \tau) .
\end{equation}
Note that $aH = -(1+\epsilon)/\tau$, and hence
\begin{equation}
 \tau = -\frac{(1+\epsilon)}{aH}  ,
\end{equation}
which is an exact relation, as it follows from the definition of $\epsilon$. Also note the asymptotic form of the Hankel function,
\begin{equation}
\lim_{k \tau \to 0} H_{\nu} ^{(1)} (- k \tau) = \frac{i}{\pi} \Gamma (\nu) \left(\frac{2}{- k \tau}\right)^{ \nu} .
\end{equation}
We plug all this into $|\phi_k|^2$, and evaluate at $k = \gamma a H$,
\begin{eqnarray}
|\phi_k|^2 &=& \frac{\Gamma(\nu)^2}{4 \pi} \frac{2^{2\nu}}{a^3 H}  {(1+\epsilon)^{1 - 2\nu}} \frac{1}{\gamma^{2\nu}}  .
\end{eqnarray}
Now we use the expression for $\nu^2$ to expand $\nu$ as:
\begin{equation}
\nu ^2 = \frac{9}{4} + \delta \rightarrow \nu \sim \frac{3}{2} + \frac{\delta}{3} ,
\end{equation}
where $\delta = 9 \epsilon - (1+2\epsilon)(m/H_0)^2$. We do a combined expansion in $(m/H)^2$ and $\epsilon$, and the variance, given in equation (\ref{appendixvariance}), simplifies to (to first order in both $\epsilon$ and $(m/H)^2$)
\begin{equation}
\langle \xi(N) \xi(N')\rangle = \frac{H^4}{4 \pi^2} \left[ 1+\Delta \right] \delta(N-N') ,
\end{equation}
where $\Delta$ is defined as
\begin{equation}
\Delta = \frac{2}{3} \left( \frac{m}{H}\right)^2 \left( - 2 + \gamma_E + \log 2 \gamma \right) - 3 \epsilon (-3 + 2 \gamma_E + 2 \log 2  \gamma) .
\end{equation}
Note that the stochastic correction is hidden inside $\epsilon$, as per equation (\ref{eq:StochasticCorrectionToEpsilon}). Also note that the $H$ appearing in the two above equations is the full $H$, i.e. stochastic and slow-roll corrected.

\renewcommand{\theequation}{B \arabic{equation}}
\renewcommand{\thefigure}{B \arabic{figure}}
\setcounter{equation}{0}  

\section{{\scshape Numerical Integration Method for Stochastic Differential Equations}} 
\normalsize

\label{NumMethod}
In order to solve equation (\ref{modefunctionStep3}) numerically, we need to modify the standard Runge-Kutta method for SDEs, in order to allow for equations with more than linear powers of the noise. The method we propose still reduces to the improved Euler method in the absence of a stochastic term.

It works as follows. 
For a system of SDEs given by
\be
	\label{standardRKSDE}
	d \vec{X} = \vec{a}(t,\vec{X})dt+\vec{b}(t,\vec{X})dW\, +\vec{c}(t,\vec{X})dW^2 \,.
\ee
with $\vec{X}$ a system of stochastic processes and $W$ the Wiener process (i.e. just regular Brownian motion). We can define the following Runge-Kutta scheme:
\bea
	&\vec{K}_1&= \,~ \vec{a}(t_n, \vec{X}_n)\delta t+ \vec{b}(t_n, \vec{X}_n)\Delta W + \vec{c}(t_n, \vec{X}_n)\Delta W ^2/\delta t \, ; \\
	&\vec{K}_2&= \,~ \vec{a}(t_{n+1}, \vec{X}_{n}+\vec{K}_1)\delta t+\vec{b}(t_{n+1}, \vec{X}_{n}+\vec{K}_1)\Delta W+ \vec{c}(t_{n+1}, \vec{X}_n+\vec{K}_1)\Delta W ^2/\delta t\, ;\\
	&\vec{X}_{n+1}&=\,~ \vec{X}_n+\frac{1}{2}\left(\vec{K}_1+\vec{K}_2 \right)\, ,
\eea
where $\Delta W$ is now a random variable sampled at each time step from a normal distribution of mean 0 and standard variation~$\sqrt{\delta t}$.

In our case, the second-order SDE that needs to be solved can be cast as a system of first-order SDEs and solved using the above scheme. The components of the vector coefficients $\vec{a}$, $\vec{b}$, and $\vec{c}$ can be read from (\ref{modefunctionStep3}).

\bibliography{StochasticBib.bib}

\begin{thebibliography}{101}
\expandafter\ifx\csname natexlab\endcsname\relax\def\natexlab#1{#1}\fi
\expandafter\ifx\csname bibnamefont\endcsname\relax
  \def\bibnamefont#1{#1}\fi
\expandafter\ifx\csname bibfnamefont\endcsname\relax
  \def\bibfnamefont#1{#1}\fi
\expandafter\ifx\csname citenamefont\endcsname\relax
  \def\citenamefont#1{#1}\fi
\expandafter\ifx\csname url\endcsname\relax
  \def\url#1{\texttt{#1}}\fi
\expandafter\ifx\csname urlprefix\endcsname\relax\def\urlprefix{URL }\fi
\providecommand{\bibinfo}[2]{#2}
\providecommand{\eprint}[2][]{\url{#2}}

\bibitem[{\citenamefont{Mukhanov and Chibisov}(1981)}]{Mukhanov:1981xt}
\bibinfo{author}{\bibfnamefont{V.~F.} \bibnamefont{Mukhanov}} \bibnamefont{and}
  \bibinfo{author}{\bibfnamefont{G.}~\bibnamefont{Chibisov}},
  \bibinfo{journal}{JETP Lett.} \textbf{\bibinfo{volume}{33}},
  \bibinfo{pages}{532} (\bibinfo{year}{1981}).

\bibitem[{\citenamefont{Mukhanov and Chibisov}(1982)}]{Mukhanov:1982nu}
\bibinfo{author}{\bibfnamefont{V.~F.} \bibnamefont{Mukhanov}} \bibnamefont{and}
  \bibinfo{author}{\bibfnamefont{G.}~\bibnamefont{Chibisov}},
  \bibinfo{journal}{Sov. Phys. JETP} \textbf{\bibinfo{volume}{56}},
  \bibinfo{pages}{258} (\bibinfo{year}{1982}).

\bibitem[{\citenamefont{Bardeen et~al.}(1983)\citenamefont{Bardeen, Steinhardt,
  and Turner}}]{Bardeen:1983qw}
\bibinfo{author}{\bibfnamefont{J.~M.} \bibnamefont{Bardeen}},
  \bibinfo{author}{\bibfnamefont{P.~J.} \bibnamefont{Steinhardt}},
  \bibnamefont{and} \bibinfo{author}{\bibfnamefont{M.~S.}
  \bibnamefont{Turner}}, \bibinfo{journal}{Phys. Rev.}
  \textbf{\bibinfo{volume}{D28}}, \bibinfo{pages}{679} (\bibinfo{year}{1983}).

\bibitem[{\citenamefont{Lyth and Riotto}(1999)}]{Lyth:1998xn}
\bibinfo{author}{\bibfnamefont{D.~H.} \bibnamefont{Lyth}} \bibnamefont{and}
  \bibinfo{author}{\bibfnamefont{A.}~\bibnamefont{Riotto}},
  \bibinfo{journal}{Phys.Rept.} \textbf{\bibinfo{volume}{314}},
  \bibinfo{pages}{1} (\bibinfo{year}{1999}), \eprint{hep-ph/9807278}.

\bibitem[{\citenamefont{Kallosh et~al.}(2014)\citenamefont{Kallosh, Linde, and
  Vercnocke}}]{Kallosh:2014vja}
\bibinfo{author}{\bibfnamefont{R.}~\bibnamefont{Kallosh}},
  \bibinfo{author}{\bibfnamefont{A.}~\bibnamefont{Linde}}, \bibnamefont{and}
  \bibinfo{author}{\bibfnamefont{B.}~\bibnamefont{Vercnocke}},
  \bibinfo{journal}{Phys.Rev.} \textbf{\bibinfo{volume}{D90}},
  \bibinfo{pages}{041303} (\bibinfo{year}{2014}), \eprint{1404.6244}.

\bibitem[{\citenamefont{Baumann and McAllister}(2014)}]{Baumann:2014nda}
\bibinfo{author}{\bibfnamefont{D.}~\bibnamefont{Baumann}} \bibnamefont{and}
  \bibinfo{author}{\bibfnamefont{L.}~\bibnamefont{McAllister}}
  (\bibinfo{year}{2014}), \eprint{1404.2601}.

\bibitem[{\citenamefont{Ade et~al.}(2014)}]{Ade:2014xna}
\bibinfo{author}{\bibfnamefont{P.}~\bibnamefont{Ade}} \bibnamefont{et~al.}
  (\bibinfo{collaboration}{BICEP2 Collaboration}),
  \bibinfo{journal}{Phys.Rev.Lett.} \textbf{\bibinfo{volume}{112}},
  \bibinfo{pages}{241101} (\bibinfo{year}{2014}), \eprint{1403.3985}.

\bibitem[{\citenamefont{et~al. (Planck~Collaboration)}(2013)}]{planck}
\bibinfo{author}{\bibfnamefont{P.~A.} \bibnamefont{et~al.
  (Planck~Collaboration)}} (\bibinfo{year}{2013}), \eprint{Planck 2013 results.
  I-XXIX. 29 Papers published in \mbox{A\&A}}.

\bibitem[{\citenamefont{Mukhanov et~al.}(1997)\citenamefont{Mukhanov, Abramo,
  and Brandenberger}}]{Mukhanov:1996ak}
\bibinfo{author}{\bibfnamefont{V.~F.} \bibnamefont{Mukhanov}},
  \bibinfo{author}{\bibfnamefont{L.~R.~W.} \bibnamefont{Abramo}},
  \bibnamefont{and} \bibinfo{author}{\bibfnamefont{R.~H.}
  \bibnamefont{Brandenberger}}, \bibinfo{journal}{Phys.Rev.Lett.}
  \textbf{\bibinfo{volume}{78}}, \bibinfo{pages}{1624} (\bibinfo{year}{1997}),
  \eprint{gr-qc/9609026}.

\bibitem[{\citenamefont{Brandenberger}(2002)}]{Brandenberger:2002sk}
\bibinfo{author}{\bibfnamefont{R.~H.} \bibnamefont{Brandenberger}}
  (\bibinfo{year}{2002}), \eprint{hep-th/0210165}.

\bibitem[{\citenamefont{Geshnizjani and Brandenberger}(2005)}]{Geshnizjani2005}
\bibinfo{author}{\bibfnamefont{G.}~\bibnamefont{Geshnizjani}} \bibnamefont{and}
  \bibinfo{author}{\bibfnamefont{R.}~\bibnamefont{Brandenberger}},
  \bibinfo{journal}{JCAP} \textbf{\bibinfo{volume}{0504}}, \bibinfo{pages}{006}
  (\bibinfo{year}{2005}), \eprint{arXiv:hep-th/0310265}.

\bibitem[{\citenamefont{Martineau and Brandenberger}(2005)}]{Martineau:2005aa}
\bibinfo{author}{\bibfnamefont{P.}~\bibnamefont{Martineau}} \bibnamefont{and}
  \bibinfo{author}{\bibfnamefont{R.~H.} \bibnamefont{Brandenberger}},
  \bibinfo{journal}{Phys.Rev.} \textbf{\bibinfo{volume}{D72}},
  \bibinfo{pages}{023507} (\bibinfo{year}{2005}), \eprint{astro-ph/0505236}.

\bibitem[{\citenamefont{{Kolb} et~al.}(2010)\citenamefont{{Kolb}, {Marra}, and
  {Matarrese}}}]{2010GReGr..42.1399K}
\bibinfo{author}{\bibfnamefont{E.~W.} \bibnamefont{{Kolb}}},
  \bibinfo{author}{\bibfnamefont{V.}~\bibnamefont{{Marra}}}, \bibnamefont{and}
  \bibinfo{author}{\bibfnamefont{S.}~\bibnamefont{{Matarrese}}},
  \bibinfo{journal}{General Relativity and Gravitation}
  \textbf{\bibinfo{volume}{42}}, \bibinfo{pages}{1399} (\bibinfo{year}{2010}),
  \eprint{0901.4566}.

\bibitem[{\citenamefont{Levasseur et~al.}(2010)\citenamefont{Levasseur,
  Laporte, and Brandenberger}}]{Levasseur2010}
\bibinfo{author}{\bibfnamefont{L.~P.} \bibnamefont{Levasseur}},
  \bibinfo{author}{\bibfnamefont{G.}~\bibnamefont{Laporte}}, \bibnamefont{and}
  \bibinfo{author}{\bibfnamefont{R.}~\bibnamefont{Brandenberger}},
  \bibinfo{journal}{Phys. Rev. D} \textbf{\bibinfo{volume}{82}},
  \bibinfo{pages}{17} (\bibinfo{year}{2010}), \eprint{arXiv:1004.1425
  [hep-th]}.

\bibitem[{\citenamefont{Burgess et~al.}(2014)\citenamefont{Burgess, Holman,
  Tasinato, and Williams}}]{Burgess:2014eoa}
\bibinfo{author}{\bibfnamefont{C.}~\bibnamefont{Burgess}},
  \bibinfo{author}{\bibfnamefont{R.}~\bibnamefont{Holman}},
  \bibinfo{author}{\bibfnamefont{G.}~\bibnamefont{Tasinato}}, \bibnamefont{and}
  \bibinfo{author}{\bibfnamefont{M.}~\bibnamefont{Williams}}
  (\bibinfo{year}{2014}), \eprint{1408.5002}.

\bibitem[{\citenamefont{Berera et~al.}(2009)\citenamefont{Berera, Moss, and
  Ramos}}]{Berera2008}
\bibinfo{author}{\bibfnamefont{A.}~\bibnamefont{Berera}},
  \bibinfo{author}{\bibfnamefont{I.~G.} \bibnamefont{Moss}}, \bibnamefont{and}
  \bibinfo{author}{\bibfnamefont{R.~O.} \bibnamefont{Ramos}},
  \bibinfo{journal}{Rept. Prog. Phys.} \textbf{\bibinfo{volume}{72}},
  \bibinfo{pages}{026901} (\bibinfo{year}{2009}), \eprint{arXiv:0808.1855
  [hep-th]}.

\bibitem[{\citenamefont{Starobinsky}(1984)}]{Starobinsky1984}
\bibinfo{author}{\bibfnamefont{A.~A.} \bibnamefont{Starobinsky}}, in
  \emph{\bibinfo{booktitle}{{Fundamental Interactions}}}, edited by
  \bibinfo{editor}{\bibfnamefont{V.~N.} \bibnamefont{Ponomarev}}
  (\bibinfo{publisher}{MGPI Press}, \bibinfo{address}{Moscow},
  \bibinfo{year}{1984}), p.~\bibinfo{pages}{54}.

\bibitem[{\citenamefont{Starobinsky}(1986)}]{Starobinsky1986}
\bibinfo{author}{\bibfnamefont{A.~A.} \bibnamefont{Starobinsky}}, in
  \emph{\bibinfo{booktitle}{Lecture Notes in Physics Vol. 246}}, edited by
  \bibinfo{editor}{\bibfnamefont{H.~J.} \bibnamefont{de~Vega}}
  \bibnamefont{and} \bibinfo{editor}{\bibfnamefont{N.}~\bibnamefont{Sanchez}}
  (\bibinfo{publisher}{Springer}, \bibinfo{address}{New York},
  \bibinfo{year}{1986}), p. \bibinfo{pages}{107}.

\bibitem[{\citenamefont{Vilenkin}(1983{\natexlab{a}})}]{Vilenkin1983a}
\bibinfo{author}{\bibfnamefont{A.}~\bibnamefont{Vilenkin}},
  \bibinfo{journal}{Phys. Rev. D} \textbf{\bibinfo{volume}{27}},
  \bibinfo{pages}{2848} (\bibinfo{year}{1983}{\natexlab{a}}).

\bibitem[{\citenamefont{Vilenkin}(1983{\natexlab{b}})}]{Vilenkin1983b}
\bibinfo{author}{\bibfnamefont{A.}~\bibnamefont{Vilenkin}},
  \bibinfo{journal}{Nucl. Phys. B} \textbf{\bibinfo{volume}{226}},
  \bibinfo{pages}{527 } (\bibinfo{year}{1983}{\natexlab{b}}).

\bibitem[{\citenamefont{Bardeen and Bublik}(1987)}]{Bardeen1986}
\bibinfo{author}{\bibfnamefont{J.~M.} \bibnamefont{Bardeen}} \bibnamefont{and}
  \bibinfo{author}{\bibfnamefont{G.~J.} \bibnamefont{Bublik}},
  \bibinfo{journal}{Class. Quant. Grav.} \textbf{\bibinfo{volume}{4}},
  \bibinfo{pages}{573} (\bibinfo{year}{1987}).

\bibitem[{\citenamefont{Rey}(1987)}]{Rey1987}
\bibinfo{author}{\bibfnamefont{S.-J.} \bibnamefont{Rey}},
  \bibinfo{journal}{Nuclear Physics B} \textbf{\bibinfo{volume}{284}},
  \bibinfo{pages}{706 } (\bibinfo{year}{1987}).

\bibitem[{\citenamefont{Goncharov et~al.}(1987)\citenamefont{Goncharov, Linde,
  and Mukhanov}}]{Goncharov:1987}
\bibinfo{author}{\bibfnamefont{A.~S.} \bibnamefont{Goncharov}},
  \bibinfo{author}{\bibfnamefont{A.~D.} \bibnamefont{Linde}}, \bibnamefont{and}
  \bibinfo{author}{\bibfnamefont{V.~F.} \bibnamefont{Mukhanov}},
  \bibinfo{journal}{International Journal of Modern Physics A}
  \textbf{\bibinfo{volume}{02}}, \bibinfo{pages}{561} (\bibinfo{year}{1987}).

\bibitem[{\citenamefont{Sasaki et~al.}(1988)\citenamefont{Sasaki, Nambu, and
  Nakao}}]{Sasaki:1987gy}
\bibinfo{author}{\bibfnamefont{M.}~\bibnamefont{Sasaki}},
  \bibinfo{author}{\bibfnamefont{Y.}~\bibnamefont{Nambu}}, \bibnamefont{and}
  \bibinfo{author}{\bibfnamefont{K.-i.} \bibnamefont{Nakao}},
  \bibinfo{journal}{Nucl. Phys. B} \textbf{\bibinfo{volume}{308}},
  \bibinfo{pages}{868} (\bibinfo{year}{1988}).

\bibitem[{\citenamefont{Nakao et~al.}(1988)\citenamefont{Nakao, Nambu, and
  Sasaki}}]{Nakao:1988yi}
\bibinfo{author}{\bibfnamefont{K.-i.} \bibnamefont{Nakao}},
  \bibinfo{author}{\bibfnamefont{Y.}~\bibnamefont{Nambu}}, \bibnamefont{and}
  \bibinfo{author}{\bibfnamefont{M.}~\bibnamefont{Sasaki}},
  \bibinfo{journal}{Prog. Theor. Phys.} \textbf{\bibinfo{volume}{80}},
  \bibinfo{pages}{1041} (\bibinfo{year}{1988}).

\bibitem[{\citenamefont{Nambu and Sasaki}(1989)}]{Nambu:1988je}
\bibinfo{author}{\bibfnamefont{Y.}~\bibnamefont{Nambu}} \bibnamefont{and}
  \bibinfo{author}{\bibfnamefont{M.}~\bibnamefont{Sasaki}},
  \bibinfo{journal}{Phys. Lett. B} \textbf{\bibinfo{volume}{219}},
  \bibinfo{pages}{240} (\bibinfo{year}{1989}).

\bibitem[{\citenamefont{Nambu}(1989)}]{Nambu:1989uf}
\bibinfo{author}{\bibfnamefont{Y.}~\bibnamefont{Nambu}},
  \bibinfo{journal}{Prog. Theor. Phys.} \textbf{\bibinfo{volume}{81}},
  \bibinfo{pages}{1037} (\bibinfo{year}{1989}).

\bibitem[{\citenamefont{Salopek and Bond}(1990)}]{Salopek1990}
\bibinfo{author}{\bibfnamefont{D.~S.} \bibnamefont{Salopek}} \bibnamefont{and}
  \bibinfo{author}{\bibfnamefont{J.~R.} \bibnamefont{Bond}},
  \bibinfo{journal}{Phys. Rev. D} \textbf{\bibinfo{volume}{42}},
  \bibinfo{pages}{3936} (\bibinfo{year}{1990}).

\bibitem[{\citenamefont{Salopek and Bond}(1991)}]{Salopek1991}
\bibinfo{author}{\bibfnamefont{D.~S.} \bibnamefont{Salopek}} \bibnamefont{and}
  \bibinfo{author}{\bibfnamefont{J.~R.} \bibnamefont{Bond}},
  \bibinfo{journal}{Phys. Rev. D} \textbf{\bibinfo{volume}{43}},
  \bibinfo{pages}{1005} (\bibinfo{year}{1991}).

\bibitem[{\citenamefont{Habib}(1992)}]{Habib1992}
\bibinfo{author}{\bibfnamefont{S.}~\bibnamefont{Habib}},
  \bibinfo{journal}{Phys. Rev. D} \textbf{\bibinfo{volume}{46}},
  \bibinfo{pages}{2408} (\bibinfo{year}{1992}).

\bibitem[{\citenamefont{Linde et~al.}(1994)\citenamefont{Linde, Linde, and
  Mezhlumian}}]{Linde:1993xx}
\bibinfo{author}{\bibfnamefont{A.~D.} \bibnamefont{Linde}},
  \bibinfo{author}{\bibfnamefont{D.~A.} \bibnamefont{Linde}}, \bibnamefont{and}
  \bibinfo{author}{\bibfnamefont{A.}~\bibnamefont{Mezhlumian}},
  \bibinfo{journal}{Phys. Rev. D} \textbf{\bibinfo{volume}{49}},
  \bibinfo{pages}{1783} (\bibinfo{year}{1994}), \eprint{arXiv:gr-qc/9306035v3}.

\bibitem[{\citenamefont{Starobinsky and Yokoyama}(1994)}]{Starobinsky1994}
\bibinfo{author}{\bibfnamefont{A.~A.} \bibnamefont{Starobinsky}}
  \bibnamefont{and} \bibinfo{author}{\bibfnamefont{J.}~\bibnamefont{Yokoyama}},
  \bibinfo{journal}{Phys. Rev. D} \textbf{\bibinfo{volume}{50}},
  \bibinfo{pages}{6357} (\bibinfo{year}{1994}),
  \eprint{arXiv:astro-ph/9407016}.

\bibitem[{\citenamefont{Morikawa}(1990)}]{Morikawa:1989xz}
\bibinfo{author}{\bibfnamefont{M.}~\bibnamefont{Morikawa}},
  \bibinfo{journal}{Phys.Rev.} \textbf{\bibinfo{volume}{D42}},
  \bibinfo{pages}{1027} (\bibinfo{year}{1990}).

\bibitem[{\citenamefont{Hosoya et~al.}(1989)\citenamefont{Hosoya, Morikawa, and
  Nakayama}}]{Hosoya:1988yz}
\bibinfo{author}{\bibfnamefont{A.}~\bibnamefont{Hosoya}},
  \bibinfo{author}{\bibfnamefont{M.}~\bibnamefont{Morikawa}}, \bibnamefont{and}
  \bibinfo{author}{\bibfnamefont{K.}~\bibnamefont{Nakayama}},
  \bibinfo{journal}{Int.J.Mod.Phys.} \textbf{\bibinfo{volume}{A4}},
  \bibinfo{pages}{2613} (\bibinfo{year}{1989}).

\bibitem[{\citenamefont{Calzetta and Hu}(1994)}]{Calzetta:1993qe}
\bibinfo{author}{\bibfnamefont{E.}~\bibnamefont{Calzetta}} \bibnamefont{and}
  \bibinfo{author}{\bibfnamefont{B.}~\bibnamefont{Hu}},
  \bibinfo{journal}{Phys.Rev.} \textbf{\bibinfo{volume}{D49}},
  \bibinfo{pages}{6636} (\bibinfo{year}{1994}), \eprint{gr-qc/9312036}.

\bibitem[{\citenamefont{Calzetta and Hu}(1995)}]{Calzetta1995}
\bibinfo{author}{\bibfnamefont{E.}~\bibnamefont{Calzetta}} \bibnamefont{and}
  \bibinfo{author}{\bibfnamefont{B.~L.} \bibnamefont{Hu}},
  \bibinfo{journal}{Phys Rev D} \textbf{\bibinfo{volume}{52}},
  \bibinfo{pages}{6770} (\bibinfo{year}{1995}), \eprint{arXiv:gr-qc/9505046}.

\bibitem[{\citenamefont{Calzetta et~al.}(1997)\citenamefont{Calzetta, Campos,
  and Verdaguer}}]{Calzetta1997}
\bibinfo{author}{\bibfnamefont{E.}~\bibnamefont{Calzetta}},
  \bibinfo{author}{\bibfnamefont{A.}~\bibnamefont{Campos}}, \bibnamefont{and}
  \bibinfo{author}{\bibfnamefont{E.}~\bibnamefont{Verdaguer}},
  \bibinfo{journal}{Phys. Rev. D} \textbf{\bibinfo{volume}{56}},
  \bibinfo{pages}{2163} (\bibinfo{year}{1997}).

\bibitem[{\citenamefont{Matacz}(1997)}]{Matacz1997}
\bibinfo{author}{\bibfnamefont{A.}~\bibnamefont{Matacz}},
  \bibinfo{journal}{Phys. Rev. D} \textbf{\bibinfo{volume}{55}},
  \bibinfo{pages}{1860} (\bibinfo{year}{1997}).

\bibitem[{\citenamefont{Calzetta and Hu}(1999)}]{Calzetta1999}
\bibinfo{author}{\bibfnamefont{E.}~\bibnamefont{Calzetta}} \bibnamefont{and}
  \bibinfo{author}{\bibfnamefont{B.}~\bibnamefont{Hu}},
  \bibinfo{journal}{Physical Review D} \textbf{\bibinfo{volume}{61}},
  \bibinfo{pages}{1} (\bibinfo{year}{1999}).

\bibitem[{\citenamefont{Matarrese et~al.}(2004)\citenamefont{Matarrese, Musso,
  and Riotto}}]{Matarrese2004}
\bibinfo{author}{\bibfnamefont{S.}~\bibnamefont{Matarrese}},
  \bibinfo{author}{\bibfnamefont{M.~A.} \bibnamefont{Musso}}, \bibnamefont{and}
  \bibinfo{author}{\bibfnamefont{A.}~\bibnamefont{Riotto}},
  \bibinfo{journal}{JCAP} \textbf{\bibinfo{volume}{0405}}, \bibinfo{pages}{008}
  (\bibinfo{year}{2004}), \eprint{arXiv:hep-th/0311059}.

\bibitem[{\citenamefont{Franco and Calzetta}(2010)}]{Franco2010}
\bibinfo{author}{\bibfnamefont{M.}~\bibnamefont{Franco}} \bibnamefont{and}
  \bibinfo{author}{\bibfnamefont{E.}~\bibnamefont{Calzetta}}
  (\bibinfo{year}{2010}), \eprint{arXiv:1002.2916 [astro-ph.CO]}.

\bibitem[{\citenamefont{{Afshordi} and {Brandenberger}}(2001)}]{Afshordi2001}
\bibinfo{author}{\bibfnamefont{N.}~\bibnamefont{{Afshordi}}} \bibnamefont{and}
  \bibinfo{author}{\bibfnamefont{R.}~\bibnamefont{{Brandenberger}}},
  \bibinfo{journal}{Phys. Rev.~D} \textbf{\bibinfo{volume}{63}},
  \bibinfo{pages}{123505} (\bibinfo{year}{2001}), \eprint{arXiv:gr-qc/0011075}.

\bibitem[{\citenamefont{Geshnizjani and
  Afshordi}(2005{\natexlab{a}})}]{Geshnizjani2005a}
\bibinfo{author}{\bibfnamefont{G.}~\bibnamefont{Geshnizjani}} \bibnamefont{and}
  \bibinfo{author}{\bibfnamefont{N.}~\bibnamefont{Afshordi}},
  \bibinfo{journal}{JCAP} \textbf{\bibinfo{volume}{0501}}, \bibinfo{pages}{011}
  (\bibinfo{year}{2005}{\natexlab{a}}), \eprint{arXiv:gr-qc/0405117v1}.

\bibitem[{\citenamefont{Tsamis and Woodard}(2005{\natexlab{a}})}]{Tsamis2005}
\bibinfo{author}{\bibfnamefont{N.}~\bibnamefont{Tsamis}} \bibnamefont{and}
  \bibinfo{author}{\bibfnamefont{R.}~\bibnamefont{Woodard}},
  \bibinfo{journal}{Nuc. Phys. B} \textbf{\bibinfo{volume}{724}},
  \bibinfo{pages}{295} (\bibinfo{year}{2005}{\natexlab{a}}), \eprint{0505115}.

\bibitem[{\citenamefont{Martin and Musso}(2006)}]{Martin2006}
\bibinfo{author}{\bibfnamefont{J.}~\bibnamefont{Martin}} \bibnamefont{and}
  \bibinfo{author}{\bibfnamefont{M.}~\bibnamefont{Musso}},
  \bibinfo{journal}{Physical Review D}  (\bibinfo{year}{2006}),
  \eprint{0511214v2}.

\bibitem[{\citenamefont{Kunze}(2006)}]{Kunze:2006tu}
\bibinfo{author}{\bibfnamefont{K.~E.} \bibnamefont{Kunze}},
  \bibinfo{journal}{JCAP} \textbf{\bibinfo{volume}{0607}}, \bibinfo{pages}{014}
  (\bibinfo{year}{2006}), \eprint{arXiv:astro-ph/0603575}.

\bibitem[{\citenamefont{Finelli et~al.}(2009)\citenamefont{Finelli, Marozzi,
  Starobinsky, Vacca, and Venturi}}]{Finelli:2008zg}
\bibinfo{author}{\bibfnamefont{F.}~\bibnamefont{Finelli}},
  \bibinfo{author}{\bibfnamefont{G.}~\bibnamefont{Marozzi}},
  \bibinfo{author}{\bibfnamefont{A.~A.} \bibnamefont{Starobinsky}},
  \bibinfo{author}{\bibfnamefont{G.~P.} \bibnamefont{Vacca}}, \bibnamefont{and}
  \bibinfo{author}{\bibfnamefont{G.}~\bibnamefont{Venturi}},
  \bibinfo{journal}{Phys. Rev. D} \textbf{\bibinfo{volume}{79}},
  \bibinfo{pages}{044007} (\bibinfo{year}{2009}), \eprint{arXiv:0808.1786
  [hep-th]}.

\bibitem[{\citenamefont{Finelli et~al.}(2010)\citenamefont{Finelli, Marozzi,
  Starobinsky, Vacca, and Venturi}}]{Finelli:2010sh}
\bibinfo{author}{\bibfnamefont{F.}~\bibnamefont{Finelli}},
  \bibinfo{author}{\bibfnamefont{G.}~\bibnamefont{Marozzi}},
  \bibinfo{author}{\bibfnamefont{A.~A.} \bibnamefont{Starobinsky}},
  \bibinfo{author}{\bibfnamefont{G.~P.} \bibnamefont{Vacca}}, \bibnamefont{and}
  \bibinfo{author}{\bibfnamefont{G.}~\bibnamefont{Venturi}},
  \bibinfo{journal}{Phys. Rev.} \textbf{\bibinfo{volume}{D82}},
  \bibinfo{pages}{064020} (\bibinfo{year}{2010}), \eprint{1003.1327}.

\bibitem[{\citenamefont{Clesse}(2011)}]{Clesse:2010iz}
\bibinfo{author}{\bibfnamefont{S.}~\bibnamefont{Clesse}},
  \bibinfo{journal}{Physical Review D} \textbf{\bibinfo{volume}{83}},
  \bibinfo{pages}{063518} (\bibinfo{year}{2011}), \eprint{arXiv:1006.4522v3}.

\bibitem[{\citenamefont{Martin and Vennin}(2012)}]{Martin2011}
\bibinfo{author}{\bibfnamefont{J.}~\bibnamefont{Martin}} \bibnamefont{and}
  \bibinfo{author}{\bibfnamefont{V.}~\bibnamefont{Vennin}},
  \bibinfo{journal}{Phys. Rev. D} \textbf{\bibinfo{volume}{85}},
  \bibinfo{pages}{043525} (\bibinfo{year}{2012}), \eprint{arXiv:1110.2070
  [astro-ph.CO]}.

\bibitem[{\citenamefont{Weenink and Prokopec}(2011)}]{Weenink2011}
\bibinfo{author}{\bibfnamefont{J.}~\bibnamefont{Weenink}} \bibnamefont{and}
  \bibinfo{author}{\bibfnamefont{T.}~\bibnamefont{Prokopec}}
  (\bibinfo{year}{2011}), \eprint{arXiv:1108.3994 [gr-qc]}.

\bibitem[{\citenamefont{Boyanovsky et~al.}(1995)\citenamefont{Boyanovsky,
  de~Vega, Holman, Lee, and Singh}}]{Boyanovsky1995}
\bibinfo{author}{\bibfnamefont{D.}~\bibnamefont{Boyanovsky}},
  \bibinfo{author}{\bibfnamefont{H.~J.} \bibnamefont{de~Vega}},
  \bibinfo{author}{\bibfnamefont{R.}~\bibnamefont{Holman}},
  \bibinfo{author}{\bibfnamefont{D.-S.} \bibnamefont{Lee}}, \bibnamefont{and}
  \bibinfo{author}{\bibfnamefont{A.}~\bibnamefont{Singh}},
  \bibinfo{journal}{Phys. Rev. D} \textbf{\bibinfo{volume}{51}},
  \bibinfo{pages}{4419} (\bibinfo{year}{1995}).

\bibitem[{\citenamefont{Weinberg}(2005)}]{Weinberg:2005vy}
\bibinfo{author}{\bibfnamefont{S.}~\bibnamefont{Weinberg}},
  \bibinfo{journal}{Phys. Rev. D} \textbf{\bibinfo{volume}{72}},
  \bibinfo{pages}{043514} (\bibinfo{year}{2005}),
  \eprint{arXiv:hep-th/0506236}.

\bibitem[{\citenamefont{Maldacena}(2003)}]{Maldacena:2002vr}
\bibinfo{author}{\bibfnamefont{J.~M.} \bibnamefont{Maldacena}},
  \bibinfo{journal}{JHEP} \textbf{\bibinfo{volume}{0305}}, \bibinfo{pages}{013}
  (\bibinfo{year}{2003}), \eprint{arXiv:astro-ph/0210603}.

\bibitem[{\citenamefont{Calzetta and Hu}(1987)}]{Calzetta1987}
\bibinfo{author}{\bibfnamefont{E.}~\bibnamefont{Calzetta}} \bibnamefont{and}
  \bibinfo{author}{\bibfnamefont{B.~L.} \bibnamefont{Hu}},
  \bibinfo{journal}{Phys. Rev. D} \textbf{\bibinfo{volume}{35}},
  \bibinfo{pages}{495} (\bibinfo{year}{1987}).

\bibitem[{\citenamefont{Jordan}(1986)}]{Jordan1986}
\bibinfo{author}{\bibfnamefont{R.~D.} \bibnamefont{Jordan}},
  \bibinfo{journal}{Phys. Rev. D} \textbf{\bibinfo{volume}{33}},
  \bibinfo{pages}{444} (\bibinfo{year}{1986}).

\bibitem[{\citenamefont{Altland and Simons}(2010)}]{Altland2010}
\bibinfo{author}{\bibfnamefont{A.}~\bibnamefont{Altland}} \bibnamefont{and}
  \bibinfo{author}{\bibfnamefont{B.}~\bibnamefont{Simons}},
  \emph{\bibinfo{title}{Condensed Matter Field Theory}}
  (\bibinfo{publisher}{Cambridge University Press}, \bibinfo{year}{2010}), ISBN
  \bibinfo{isbn}{9780521769754}.

\bibitem[{\citenamefont{Calzetta and Hu}(2008)}]{Calzetta2008}
\bibinfo{author}{\bibfnamefont{E.}~\bibnamefont{Calzetta}} \bibnamefont{and}
  \bibinfo{author}{\bibfnamefont{B.}~\bibnamefont{Hu}},
  \emph{\bibinfo{title}{Nonequilibrium quantum field theory}}
  (\bibinfo{publisher}{Cambridge University Press}, \bibinfo{year}{2008}), ISBN
  \bibinfo{isbn}{9780521641685}.

\bibitem[{\citenamefont{Perreault~Levasseur}(2013)}]{Levasseur:2013ffa}
\bibinfo{author}{\bibfnamefont{L.}~\bibnamefont{Perreault~Levasseur}},
  \bibinfo{journal}{PHYS. REV. D88,} \textbf{\bibinfo{volume}{083537}},
  \bibinfo{pages}{083537} (\bibinfo{year}{2013}), \eprint{1304.6408}.

\bibitem[{\citenamefont{Fujita et~al.}(2013)\citenamefont{Fujita, Kawasaki,
  Tada, and Takesako}}]{Fujita:2013cna}
\bibinfo{author}{\bibfnamefont{T.}~\bibnamefont{Fujita}},
  \bibinfo{author}{\bibfnamefont{M.}~\bibnamefont{Kawasaki}},
  \bibinfo{author}{\bibfnamefont{Y.}~\bibnamefont{Tada}}, \bibnamefont{and}
  \bibinfo{author}{\bibfnamefont{T.}~\bibnamefont{Takesako}},
  \bibinfo{journal}{JCAP} \textbf{\bibinfo{volume}{1312}}, \bibinfo{pages}{036}
  (\bibinfo{year}{2013}), \eprint{1308.4754}.

\bibitem[{\citenamefont{Fujita et~al.}(2014)\citenamefont{Fujita, Kawasaki, and
  Tada}}]{Fujita:2014tja}
\bibinfo{author}{\bibfnamefont{T.}~\bibnamefont{Fujita}},
  \bibinfo{author}{\bibfnamefont{M.}~\bibnamefont{Kawasaki}}, \bibnamefont{and}
  \bibinfo{author}{\bibfnamefont{Y.}~\bibnamefont{Tada}}
  (\bibinfo{year}{2014}), \eprint{1405.2187}.

\bibitem[{\citenamefont{Perreault~Levasseur
  et~al.}(2013)\citenamefont{Perreault~Levasseur, Vennin, and
  Brandenberger}}]{Levasseur:2013tja}
\bibinfo{author}{\bibfnamefont{L.}~\bibnamefont{Perreault~Levasseur}},
  \bibinfo{author}{\bibfnamefont{V.}~\bibnamefont{Vennin}}, \bibnamefont{and}
  \bibinfo{author}{\bibfnamefont{R.}~\bibnamefont{Brandenberger}},
  \bibinfo{journal}{Phys.Rev.} \textbf{\bibinfo{volume}{D88}},
  \bibinfo{pages}{083538} (\bibinfo{year}{2013}), \eprint{1307.2575}.

\bibitem[{\citenamefont{Linde}(1994{\natexlab{a}})}]{Linde:1993cn}
\bibinfo{author}{\bibfnamefont{A.~D.} \bibnamefont{Linde}},
  \bibinfo{journal}{Phys. Rev.} \textbf{\bibinfo{volume}{D49}},
  \bibinfo{pages}{748} (\bibinfo{year}{1994}{\natexlab{a}}),
  \eprint{astro-ph/9307002}.

\bibitem[{\citenamefont{Clesse and Rocher}(2009)}]{Clesse2009}
\bibinfo{author}{\bibfnamefont{S.}~\bibnamefont{Clesse}} \bibnamefont{and}
  \bibinfo{author}{\bibfnamefont{J.}~\bibnamefont{Rocher}},
  \bibinfo{journal}{Physical Review D} \textbf{\bibinfo{volume}{79}},
  \bibinfo{pages}{103507} (\bibinfo{year}{2009}), \eprint{arXiv:0809.4355v2}.

\bibitem[{\citenamefont{Linde}(1983)}]{Linde1983177}
\bibinfo{author}{\bibfnamefont{A.}~\bibnamefont{Linde}},
  \bibinfo{journal}{Phys. Lett. B} \textbf{\bibinfo{volume}{129}},
  \bibinfo{pages}{177 } (\bibinfo{year}{1983}).

\bibitem[{\citenamefont{Linde}(1994{\natexlab{b}})}]{Linde:1994yf}
\bibinfo{author}{\bibfnamefont{A.~D.} \bibnamefont{Linde}},
  \bibinfo{journal}{{\it Lectures on Inflationary Cosmology}, Stanford preprint
  SU-ITP-94-36}  (\bibinfo{year}{1994}{\natexlab{b}}), \eprint{hep-th/9410082}.

\bibitem[{\citenamefont{Linde}(1986{\natexlab{a}})}]{Linde1986eternalinflation}
\bibinfo{author}{\bibfnamefont{A.}~\bibnamefont{Linde}},
  \bibinfo{journal}{Modern Physics Letters A} \textbf{\bibinfo{volume}{01}},
  \bibinfo{pages}{81} (\bibinfo{year}{1986}{\natexlab{a}}).

\bibitem[{\citenamefont{Linde}(1986{\natexlab{b}})}]{Linde1986395}
\bibinfo{author}{\bibfnamefont{A.}~\bibnamefont{Linde}},
  \bibinfo{journal}{Physics Letters B} \textbf{\bibinfo{volume}{175}},
  \bibinfo{pages}{395 } (\bibinfo{year}{1986}{\natexlab{b}}).

\bibitem[{\citenamefont{Guth}(2007)}]{Guth:2007ng}
\bibinfo{author}{\bibfnamefont{A.~H.} \bibnamefont{Guth}},
  \bibinfo{journal}{J.Phys.} \textbf{\bibinfo{volume}{A40}},
  \bibinfo{pages}{6811} (\bibinfo{year}{2007}), \eprint{hep-th/0702178}.

\bibitem[{\citenamefont{Li and Wang}(2007)}]{Li:2007uc}
\bibinfo{author}{\bibfnamefont{M.}~\bibnamefont{Li}} \bibnamefont{and}
  \bibinfo{author}{\bibfnamefont{Y.}~\bibnamefont{Wang}},
  \bibinfo{journal}{JCAP} \textbf{\bibinfo{volume}{0708}}, \bibinfo{pages}{007}
  (\bibinfo{year}{2007}), \eprint{0706.1691}.

\bibitem[{\citenamefont{Kohli and Haslam}(2014)}]{Kohli:2014ala}
\bibinfo{author}{\bibfnamefont{I.~S.} \bibnamefont{Kohli}} \bibnamefont{and}
  \bibinfo{author}{\bibfnamefont{M.~C.} \bibnamefont{Haslam}}
  (\bibinfo{year}{2014}), \eprint{1408.2249}.

\bibitem[{\citenamefont{Feng et~al.}(2010)\citenamefont{Feng, Li, and
  Saridakis}}]{Feng:2010ya}
\bibinfo{author}{\bibfnamefont{C.-J.} \bibnamefont{Feng}},
  \bibinfo{author}{\bibfnamefont{X.-Z.} \bibnamefont{Li}}, \bibnamefont{and}
  \bibinfo{author}{\bibfnamefont{E.~N.} \bibnamefont{Saridakis}},
  \bibinfo{journal}{Phys.Rev.} \textbf{\bibinfo{volume}{D82}},
  \bibinfo{pages}{023526} (\bibinfo{year}{2010}), \eprint{1004.1874}.

\bibitem[{\citenamefont{Qiu and Saridakis}(2012)}]{Qiu:2011zr}
\bibinfo{author}{\bibfnamefont{T.}~\bibnamefont{Qiu}} \bibnamefont{and}
  \bibinfo{author}{\bibfnamefont{E.~N.} \bibnamefont{Saridakis}},
  \bibinfo{journal}{Phys.Rev.} \textbf{\bibinfo{volume}{D85}},
  \bibinfo{pages}{043504} (\bibinfo{year}{2012}), \eprint{1107.1013}.

\bibitem[{\citenamefont{Albrecht et~al.}(1994)\citenamefont{Albrecht, Ferreira,
  Joyce, and Prokopec}}]{Albrecht1992}
\bibinfo{author}{\bibfnamefont{A.}~\bibnamefont{Albrecht}},
  \bibinfo{author}{\bibfnamefont{P.}~\bibnamefont{Ferreira}},
  \bibinfo{author}{\bibfnamefont{M.}~\bibnamefont{Joyce}}, \bibnamefont{and}
  \bibinfo{author}{\bibfnamefont{T.}~\bibnamefont{Prokopec}},
  \bibinfo{journal}{Phys. Rev. D} \textbf{\bibinfo{volume}{50}},
  \bibinfo{pages}{4807} (\bibinfo{year}{1994}),
  \eprint{arXiv:astro-ph/9303001}.

\bibitem[{\citenamefont{Hu and Sinha}(1995)}]{Hu1995}
\bibinfo{author}{\bibfnamefont{B.~L.} \bibnamefont{Hu}} \bibnamefont{and}
  \bibinfo{author}{\bibfnamefont{S.}~\bibnamefont{Sinha}},
  \bibinfo{journal}{Phys. Rev. D} \textbf{\bibinfo{volume}{51}}
  (\bibinfo{year}{1995}), \eprint{arXiv:gr-qc/9403054}.

\bibitem[{\citenamefont{Polarski and Starobinsky}(1996)}]{Polarski1995}
\bibinfo{author}{\bibfnamefont{D.}~\bibnamefont{Polarski}} \bibnamefont{and}
  \bibinfo{author}{\bibfnamefont{A.~A.} \bibnamefont{Starobinsky}},
  \bibinfo{journal}{Class. Quant. Grav.} \textbf{\bibinfo{volume}{13}},
  \bibinfo{pages}{377} (\bibinfo{year}{1996}), \eprint{arXiv:gr-qc/9504030}.

\bibitem[{\citenamefont{Kiefer et~al.}(1998)\citenamefont{Kiefer, Polarski, and
  Starobinsky}}]{Kiefer1998}
\bibinfo{author}{\bibfnamefont{C.}~\bibnamefont{Kiefer}},
  \bibinfo{author}{\bibfnamefont{D.}~\bibnamefont{Polarski}}, \bibnamefont{and}
  \bibinfo{author}{\bibfnamefont{A.~A.} \bibnamefont{Starobinsky}},
  \bibinfo{journal}{Int. J. Mod. Phys. D} \textbf{\bibinfo{volume}{7}},
  \bibinfo{pages}{455} (\bibinfo{year}{1998}), \eprint{arXiv:gr-qc/9802003}.

\bibitem[{\citenamefont{Kiefer et~al.}(2007{\natexlab{a}})\citenamefont{Kiefer,
  Lohmar, Polarski, and Starobinsky}}]{Kiefer2006}
\bibinfo{author}{\bibfnamefont{C.}~\bibnamefont{Kiefer}},
  \bibinfo{author}{\bibfnamefont{I.}~\bibnamefont{Lohmar}},
  \bibinfo{author}{\bibfnamefont{D.}~\bibnamefont{Polarski}}, \bibnamefont{and}
  \bibinfo{author}{\bibfnamefont{A.~A.} \bibnamefont{Starobinsky}},
  \bibinfo{journal}{Class. Quant. Grav.} \textbf{\bibinfo{volume}{24}},
  \bibinfo{pages}{1699} (\bibinfo{year}{2007}{\natexlab{a}}),
  \eprint{arXiv:astro-ph/0610700}.

\bibitem[{\citenamefont{Kiefer et~al.}(2007{\natexlab{b}})\citenamefont{Kiefer,
  Lohmar, Polarski, and Starobinsky}}]{Kiefer2007}
\bibinfo{author}{\bibfnamefont{C.}~\bibnamefont{Kiefer}},
  \bibinfo{author}{\bibfnamefont{I.}~\bibnamefont{Lohmar}},
  \bibinfo{author}{\bibfnamefont{D.}~\bibnamefont{Polarski}}, \bibnamefont{and}
  \bibinfo{author}{\bibfnamefont{A.~A.} \bibnamefont{Starobinsky}},
  \bibinfo{journal}{J. Phys. Conf. Ser.} \textbf{\bibinfo{volume}{67}},
  \bibinfo{pages}{012023} (\bibinfo{year}{2007}{\natexlab{b}}).

\bibitem[{\citenamefont{Tsamis and
  Woodard}(2005{\natexlab{b}})}]{Tsamis:2005hd}
\bibinfo{author}{\bibfnamefont{N.}~\bibnamefont{Tsamis}} \bibnamefont{and}
  \bibinfo{author}{\bibfnamefont{R.}~\bibnamefont{Woodard}},
  \bibinfo{journal}{Nucl.Phys.} \textbf{\bibinfo{volume}{B724}},
  \bibinfo{pages}{295} (\bibinfo{year}{2005}{\natexlab{b}}),
  \eprint{gr-qc/0505115}.

\bibitem[{\citenamefont{Casini et~al.}(1999)\citenamefont{Casini, Montemayor,
  and Sisterna}}]{Casini1999}
\bibinfo{author}{\bibfnamefont{H.}~\bibnamefont{Casini}},
  \bibinfo{author}{\bibfnamefont{R.}~\bibnamefont{Montemayor}},
  \bibnamefont{and} \bibinfo{author}{\bibfnamefont{P.}~\bibnamefont{Sisterna}},
  \bibinfo{journal}{Phys. Rev. D} \textbf{\bibinfo{volume}{59}},
  \bibinfo{pages}{063512} (\bibinfo{year}{1999}), \eprint{arXiv:gr-qc/9811083}.

\bibitem[{\citenamefont{Winitzki and Vilenkin}(2000)}]{Winitzki2000}
\bibinfo{author}{\bibfnamefont{S.}~\bibnamefont{Winitzki}} \bibnamefont{and}
  \bibinfo{author}{\bibfnamefont{A.}~\bibnamefont{Vilenkin}},
  \bibinfo{journal}{Phys. Rev. D} \textbf{\bibinfo{volume}{61}},
  \bibinfo{pages}{084008} (\bibinfo{year}{2000}), \eprint{arXiv:gr-qc/9911029}.

\bibitem[{\citenamefont{Liguori et~al.}(2004)\citenamefont{Liguori, Matarrese,
  Musso, and Riotto}}]{Liguori2004}
\bibinfo{author}{\bibfnamefont{M.}~\bibnamefont{Liguori}},
  \bibinfo{author}{\bibfnamefont{S.}~\bibnamefont{Matarrese}},
  \bibinfo{author}{\bibfnamefont{M.}~\bibnamefont{Musso}}, \bibnamefont{and}
  \bibinfo{author}{\bibfnamefont{A.}~\bibnamefont{Riotto}},
  \bibinfo{journal}{JCAP} \textbf{\bibinfo{volume}{0408}}, \bibinfo{pages}{011}
  (\bibinfo{year}{2004}), \eprint{astro-ph/0405544}.

\bibitem[{\citenamefont{Geshnizjani and
  Afshordi}(2005{\natexlab{b}})}]{Geshnizjani:2004tf}
\bibinfo{author}{\bibfnamefont{G.}~\bibnamefont{Geshnizjani}} \bibnamefont{and}
  \bibinfo{author}{\bibfnamefont{N.}~\bibnamefont{Afshordi}},
  \bibinfo{journal}{JCAP} \textbf{\bibinfo{volume}{0501}}, \bibinfo{pages}{011}
  (\bibinfo{year}{2005}{\natexlab{b}}), \eprint{gr-qc/0405117}.

\bibitem[{\citenamefont{Gordon et~al.}(2001)\citenamefont{Gordon, Wands,
  Bassett, and Maartens}}]{Gordon:2000hv}
\bibinfo{author}{\bibfnamefont{C.}~\bibnamefont{Gordon}},
  \bibinfo{author}{\bibfnamefont{D.}~\bibnamefont{Wands}},
  \bibinfo{author}{\bibfnamefont{B.~A.} \bibnamefont{Bassett}},
  \bibnamefont{and} \bibinfo{author}{\bibfnamefont{R.}~\bibnamefont{Maartens}},
  \bibinfo{journal}{Phys.Rev.} \textbf{\bibinfo{volume}{D63}},
  \bibinfo{pages}{023506} (\bibinfo{year}{2001}), \eprint{astro-ph/0009131}.

\bibitem[{\citenamefont{Sugiyama et~al.}(2013)\citenamefont{Sugiyama, Komatsu,
  and Futamase}}]{Sugiyama:2012tj}
\bibinfo{author}{\bibfnamefont{N.~S.} \bibnamefont{Sugiyama}},
  \bibinfo{author}{\bibfnamefont{E.}~\bibnamefont{Komatsu}}, \bibnamefont{and}
  \bibinfo{author}{\bibfnamefont{T.}~\bibnamefont{Futamase}},
  \bibinfo{journal}{Phys.Rev.} \textbf{\bibinfo{volume}{D87}},
  \bibinfo{pages}{023530} (\bibinfo{year}{2013}), \eprint{1208.1073}.

\bibitem[{\citenamefont{Tolley and Wyman}(2008)}]{Tolley:2008na}
\bibinfo{author}{\bibfnamefont{A.~J.} \bibnamefont{Tolley}} \bibnamefont{and}
  \bibinfo{author}{\bibfnamefont{M.}~\bibnamefont{Wyman}},
  \bibinfo{journal}{JCAP} \textbf{\bibinfo{volume}{0804}}, \bibinfo{pages}{028}
  (\bibinfo{year}{2008}), \eprint{0801.1854}.

\bibitem[{\citenamefont{Lyth and Liddle}(2009)}]{textbook}
\bibinfo{author}{\bibfnamefont{D.~H.} \bibnamefont{Lyth}} \bibnamefont{and}
  \bibinfo{author}{\bibfnamefont{A.~R.} \bibnamefont{Liddle}},
  \emph{\bibinfo{title}{{The primordial density perturbation: Cosmology,
  inflation and the origin of structure}}} (\bibinfo{year}{2009}).

\bibitem[{\citenamefont{Martin}(2004)}]{Martin:2003bt}
\bibinfo{author}{\bibfnamefont{J.}~\bibnamefont{Martin}},
  \bibinfo{journal}{Braz.J.Phys.} \textbf{\bibinfo{volume}{34}},
  \bibinfo{pages}{1307} (\bibinfo{year}{2004}), \eprint{astro-ph/0312492}.

\bibitem[{\citenamefont{Finelli et~al.}(2004)\citenamefont{Finelli, Marozzi,
  Vacca, and Venturi}}]{Finelli2003}
\bibinfo{author}{\bibfnamefont{F.}~\bibnamefont{Finelli}},
  \bibinfo{author}{\bibfnamefont{G.}~\bibnamefont{Marozzi}},
  \bibinfo{author}{\bibfnamefont{G.~P.} \bibnamefont{Vacca}}, \bibnamefont{and}
  \bibinfo{author}{\bibfnamefont{G.}~\bibnamefont{Venturi}},
  \bibinfo{journal}{Phys.Rev.} \textbf{\bibinfo{volume}{D69}},
  \bibinfo{pages}{123508} (\bibinfo{year}{2004}), \eprint{arXiv:gr-qc/0310086}.

\bibitem[{\citenamefont{Stewart and Lyth}(1993)}]{Stewart:1993bc}
\bibinfo{author}{\bibfnamefont{E.~D.} \bibnamefont{Stewart}} \bibnamefont{and}
  \bibinfo{author}{\bibfnamefont{D.~H.} \bibnamefont{Lyth}},
  \bibinfo{journal}{Phys.Lett.} \textbf{\bibinfo{volume}{B302}},
  \bibinfo{pages}{171} (\bibinfo{year}{1993}), \eprint{gr-qc/9302019}.

\bibitem[{\citenamefont{Casadio et~al.}(2005)\citenamefont{Casadio, Finelli,
  Luzzi, and Venturi}}]{Casadio:2005xv}
\bibinfo{author}{\bibfnamefont{R.}~\bibnamefont{Casadio}},
  \bibinfo{author}{\bibfnamefont{F.}~\bibnamefont{Finelli}},
  \bibinfo{author}{\bibfnamefont{M.}~\bibnamefont{Luzzi}}, \bibnamefont{and}
  \bibinfo{author}{\bibfnamefont{G.}~\bibnamefont{Venturi}},
  \bibinfo{journal}{Phys.Lett.} \textbf{\bibinfo{volume}{B625}},
  \bibinfo{pages}{1} (\bibinfo{year}{2005}), \eprint{gr-qc/0506043}.

\bibitem[{\citenamefont{Chen and Wang}(2010{\natexlab{a}})}]{Chen:2009we}
\bibinfo{author}{\bibfnamefont{X.}~\bibnamefont{Chen}} \bibnamefont{and}
  \bibinfo{author}{\bibfnamefont{Y.}~\bibnamefont{Wang}},
  \bibinfo{journal}{Phys.Rev.} \textbf{\bibinfo{volume}{D81}},
  \bibinfo{pages}{063511} (\bibinfo{year}{2010}{\natexlab{a}}),
  \eprint{0909.0496}.

\bibitem[{\citenamefont{Chen and Wang}(2010{\natexlab{b}})}]{Chen:2009zp}
\bibinfo{author}{\bibfnamefont{X.}~\bibnamefont{Chen}} \bibnamefont{and}
  \bibinfo{author}{\bibfnamefont{Y.}~\bibnamefont{Wang}},
  \bibinfo{journal}{JCAP} \textbf{\bibinfo{volume}{1004}}, \bibinfo{pages}{027}
  (\bibinfo{year}{2010}{\natexlab{b}}), \eprint{0911.3380}.

\bibitem[{\citenamefont{Achucarro
  et~al.}(2012{\natexlab{a}})\citenamefont{Achucarro, Gong, Hardeman, Palma,
  and Patil}}]{Achucarro:2012sm}
\bibinfo{author}{\bibfnamefont{A.}~\bibnamefont{Achucarro}},
  \bibinfo{author}{\bibfnamefont{J.-O.} \bibnamefont{Gong}},
  \bibinfo{author}{\bibfnamefont{S.}~\bibnamefont{Hardeman}},
  \bibinfo{author}{\bibfnamefont{G.~A.} \bibnamefont{Palma}}, \bibnamefont{and}
  \bibinfo{author}{\bibfnamefont{S.~P.} \bibnamefont{Patil}},
  \bibinfo{journal}{JHEP} \textbf{\bibinfo{volume}{1205}}, \bibinfo{pages}{066}
  (\bibinfo{year}{2012}{\natexlab{a}}), \eprint{1201.6342}.

\bibitem[{\citenamefont{Achucarro
  et~al.}(2012{\natexlab{b}})\citenamefont{Achucarro, Atal, Cespedes, Gong,
  Palma et~al.}}]{Achucarro:2012yr}
\bibinfo{author}{\bibfnamefont{A.}~\bibnamefont{Achucarro}},
  \bibinfo{author}{\bibfnamefont{V.}~\bibnamefont{Atal}},
  \bibinfo{author}{\bibfnamefont{S.}~\bibnamefont{Cespedes}},
  \bibinfo{author}{\bibfnamefont{J.-O.} \bibnamefont{Gong}},
  \bibinfo{author}{\bibfnamefont{G.~A.} \bibnamefont{Palma}},
  \bibnamefont{et~al.}, \bibinfo{journal}{Phys.Rev.}
  \textbf{\bibinfo{volume}{D86}}, \bibinfo{pages}{121301}
  (\bibinfo{year}{2012}{\natexlab{b}}), \eprint{1205.0710}.

\bibitem[{\citenamefont{Gao et~al.}(2013)\citenamefont{Gao, Langlois, and
  Mizuno}}]{Gao:2013ota}
\bibinfo{author}{\bibfnamefont{X.}~\bibnamefont{Gao}},
  \bibinfo{author}{\bibfnamefont{D.}~\bibnamefont{Langlois}}, \bibnamefont{and}
  \bibinfo{author}{\bibfnamefont{S.}~\bibnamefont{Mizuno}}
  (\bibinfo{year}{2013}), \eprint{1306.5680}.

\bibitem[{\citenamefont{Perreault~Levasseur}(2015)}]{InPrep}
\bibinfo{author}{\bibfnamefont{L.}~\bibnamefont{Perreault~Levasseur}},
  \bibinfo{journal}{In prep.}  (\bibinfo{year}{2015}).

\bibitem[{\citenamefont{Acquaviva et~al.}(2003)\citenamefont{Acquaviva,
  Bartolo, Matarrese, and Riotto}}]{Acquaviva:2002ud}
\bibinfo{author}{\bibfnamefont{V.}~\bibnamefont{Acquaviva}},
  \bibinfo{author}{\bibfnamefont{N.}~\bibnamefont{Bartolo}},
  \bibinfo{author}{\bibfnamefont{S.}~\bibnamefont{Matarrese}},
  \bibnamefont{and} \bibinfo{author}{\bibfnamefont{A.}~\bibnamefont{Riotto}},
  \bibinfo{journal}{Nucl.Phys.} \textbf{\bibinfo{volume}{B667}},
  \bibinfo{pages}{119} (\bibinfo{year}{2003}), \eprint{astro-ph/0209156}.

\bibitem[{\citenamefont{{Gardiner}}(1984)}]{Gardiner1984}
\bibinfo{author}{\bibfnamefont{C.~W.} \bibnamefont{{Gardiner}}},
  \emph{\bibinfo{title}{{Handbook of stochastic methods volume 13 of the
  Springer series in synergetics}}}, vol.~\bibinfo{volume}{1}
  (\bibinfo{year}{1984}).

\bibitem[{\citenamefont{{Volpe} et~al.}(2010)\citenamefont{{Volpe}, {Helden},
  {Brettschneider}, {Wehr}, and {Bechinger}}}]{Volpe2010}
\bibinfo{author}{\bibfnamefont{G.}~\bibnamefont{{Volpe}}},
  \bibinfo{author}{\bibfnamefont{L.}~\bibnamefont{{Helden}}},
  \bibinfo{author}{\bibfnamefont{T.}~\bibnamefont{{Brettschneider}}},
  \bibinfo{author}{\bibfnamefont{J.}~\bibnamefont{{Wehr}}}, \bibnamefont{and}
  \bibinfo{author}{\bibfnamefont{C.}~\bibnamefont{{Bechinger}}},
  \bibinfo{journal}{Phys. Rev. Lett.} \textbf{\bibinfo{volume}{104}},
  \bibinfo{eid}{170602} (\bibinfo{year}{2010}), \eprint{1004.0874}.

\end{thebibliography}

\end{document}